\documentclass[pra,amssymb,amsmath,showpacs]{revtex4}

\usepackage{graphicx}
\usepackage{amssymb}
\usepackage{amsmath}

\newcommand{\be}{\begin{equation}}
\newcommand{\ee}{\end{equation}}
\newcommand{\bea}{\begin{eqnarray}}
\newcommand{\eea}{\end{eqnarray}}

\begin{document}

\title{Test of Quantum Action for Inverse Square Potential}

\author{D.~Huard$^{a}$, H.~Kr\"{o}ger$^{a}$$\footnote{Corresponding author, Email: hkroger@phy.ulaval.ca}$,  
G.~Melkonyan$^{a}$, K.J.M.~ Moriarty$^{b}$ 
and L.P.~Nadeau$^{a}$} 

\affiliation{
$^{a}$ {\small\sl D\'{e}partement de Physique, Universit\'{e} Laval, Qu\'{e}bec, Qu\'{e}bec G1K 7P4, Canada} \\ 
$^{b}$ {\small\sl Department of Mathematics, Statistics 
and Computer Science, Dalhousie University, Halifax N.S. B3H 3J5, Canada} 
}

\begin{abstract}
We present a numerical study of the quantum action previously introduced as a parametrisation of Q.M. transition amplitudes. 
We address the questions: Is the quantum action possibly an exact parametrisation in the whole range of transition times ($0 < T < \infty$)?
Is the presence of potential terms  beyond those occuring in the classical potential required? What is the error of the parametrisation estimated from the numerical fit? How about convergence and stability of the fitting method
(dependence on grid points, resolution, initial conditions, 
internal precision etc.)?
Further we compare two methods of numerical determination of the quantum action: (i) global fit of the Q.M. transition amplitudes and (ii) flow equation.
As model we consider the inverse square potential, for which the Q.M. transition amplitudes are analytically known. 
We find that the relative error of the parametrisation starts from zero at $T=0$ increases to about $10^{-3}$ at $T=1/E_{gr}$ and then decreases to zero when $T \to \infty$. Second, we observe stability of the quantum action under variation of the control parameters. 
Finally, the flow equation method works well in the regime of large $T$ 
giving stable results under variation of initial data and
consistent with the global fit method.   
\end{abstract}


\maketitle

\setcounter{page}{0}

\newpage

\section{What is the quantum action?} 
\label{sec:WhatIs}

In Refs.\cite{Q1}-\cite{Q5} the quantum action has been introduced.
It concerns the foundations of quantum physics. It basically states that transition amplitudes in quantum mechanics can be expressed in terms of an action, which has the form of the classical action, but with parameters (mass, potential) which are different from those of the classical action. 
This presents a new link between quantum physics and classical physics. 
The quantum action can be interpreted as a renormalized action in Q.M. \cite{Q5}.
The quantum action is similar to the standard effective action, 
however, it is free of the deseases of the latter (infinite series of higher derivative terms, non-localities etc.). 
In the limit of large imaginary transition time, the existence of the quantum action has been proven \cite{Q4}. 
In this limit, the quantum action has a number of remarkable properties: 
(i) The WKB approximation \cite{WKB} for the ground state wave function becomes exact, after replacing the classical action by the quantum action.  
(ii) There is a differential equation relating the classical mass and potential to the quantum mass and potential.
(iii) There is an analytic expression for the ground state wave function in terms of the quantum action.
(iv) The ground state energy coincides with the minimum of the quantum potential. 
(v) The ground state wave function has a maximum exactly at the same position, where the quantum potential has a minimum. 
(vi) The quantum action allows also to reproduce energies and wave functions of excited states. Example: The spectrum of the hydrogen atom, considering the lowest energy states for given angular momentum.

\begin{figure}[th]
\vspace{9pt}
\begin{center}
\includegraphics[scale=0.4,angle=0]{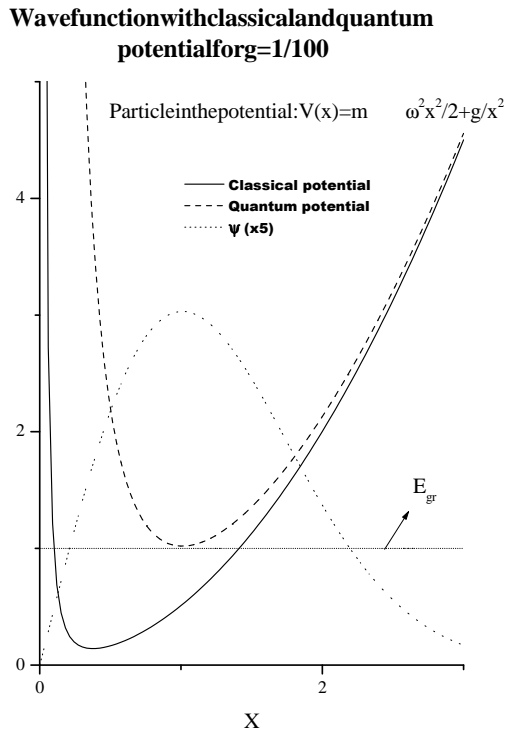}
\end{center}
\caption{Inverse square potential. Classical potential with parameters $m=1$, $\omega=1$ and $g=10^{-2}$ (full line), quantum potential (dashed line) and ground state wave function (enhanced by factor 5, dotted line). }
\label{Fig_PotWave}
\end{figure}

\begin{figure}[thb]
\vspace{9pt}
\begin{center}
\includegraphics[scale=0.4,angle=270]{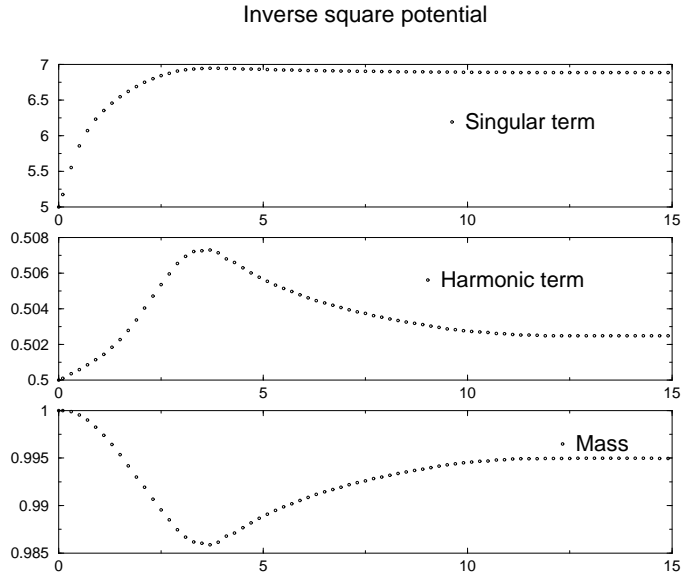}
\end{center}
\caption{Inverse square potential. Parameters of the quantum action 
$\tilde{m}$ (bottom), $\tilde{v}_{2}$ (center) and $\tilde{v}_{-2}$ (top) as function of transition time $T$. 
Classical parameters $m=1$, $v_{2}=0.5$ and $v_{-2}=5$. }
\label{Fig_ActPar}
\end{figure}

\begin{figure}[thb]
\vspace{9pt}
\begin{center}
\includegraphics[scale=0.4,angle=270]{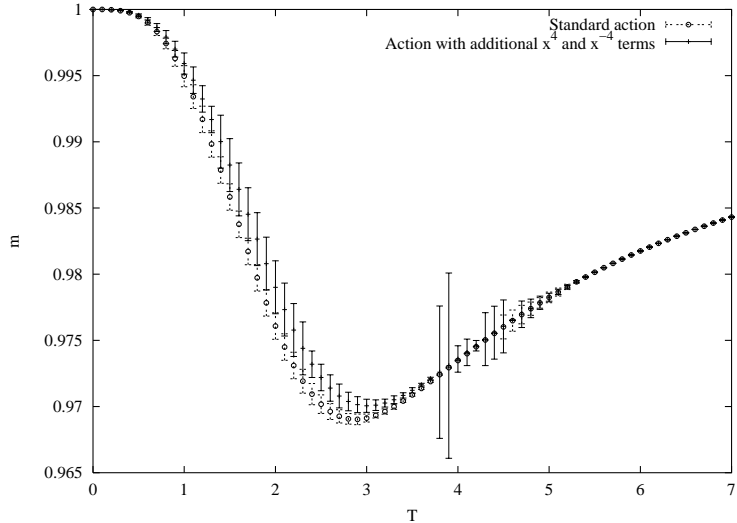}
\end{center}
\caption{Inverse square potential. Parameter $\tilde{m}$ of quantum action vs. transition time $T$. Quantum action is parametrized by supplementary terms $x^{4}$ and $x^{-4}$. Classical parameters $m=1$, $v_{2}=0.5$ ($\omega=1$), $v_{-2}=1$ ($g=1$).
Boundary points of transition: $x_{i}$ - 2 points in interval $[4,5]$; 
$x_{f}$ - 10 points in interval $[0.5,3]$. Temporal resolution 
$\Delta t = 2 ~ 10^{-3}$. }
\label{Fig_e44_m}
\end{figure}

\begin{figure}[thb]
\vspace{9pt}
\begin{center}
\includegraphics[scale=0.4,angle=270]{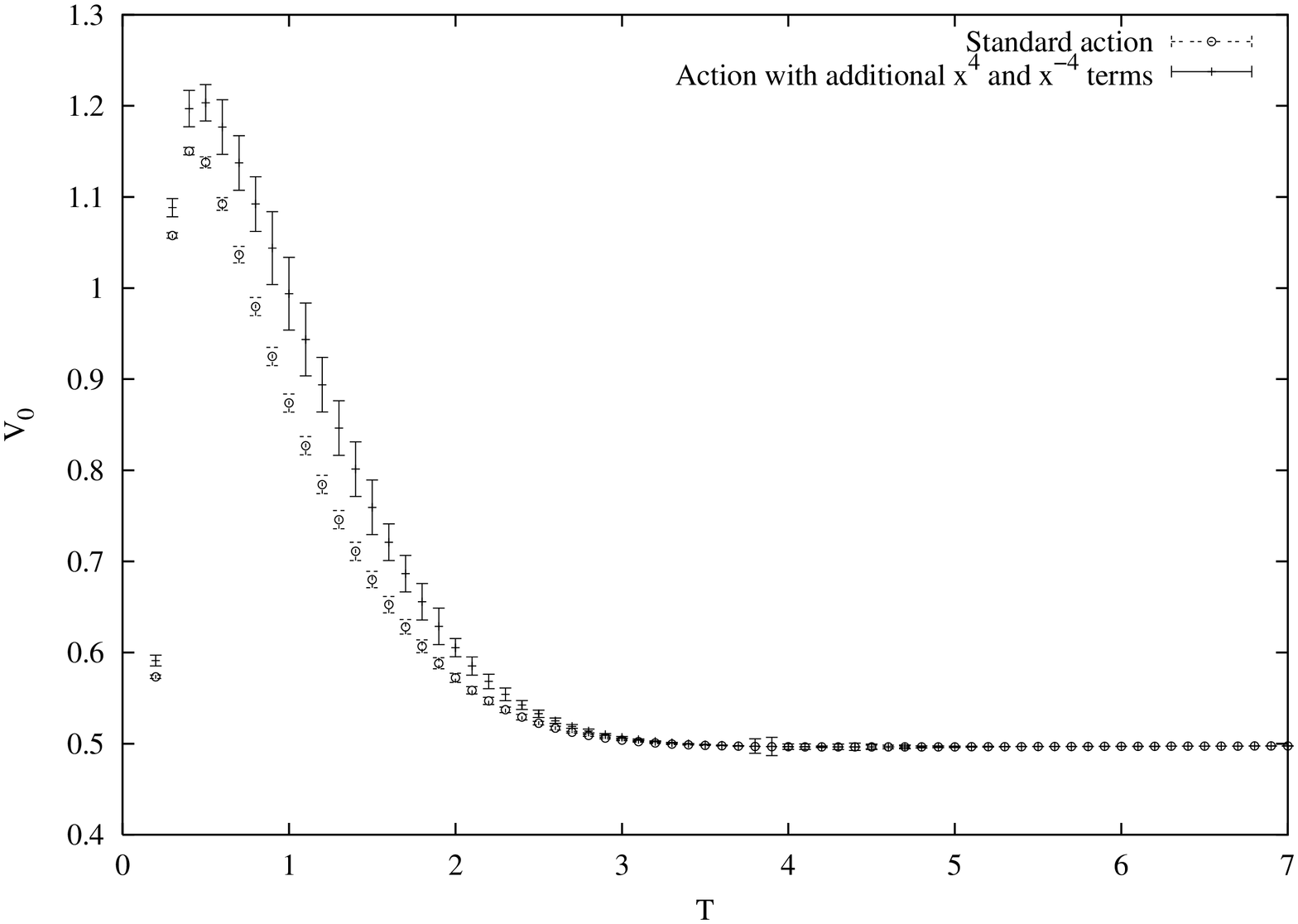}
\end{center}
\caption{Same as Fig.[\ref{Fig_e44_m}] for parameter $\tilde{v}_{0}$. }
\label{Fig_e44_v0}
\end{figure}

\begin{figure}[thb]
\vspace{9pt}
\begin{center}
\includegraphics[scale=0.4,angle=270]{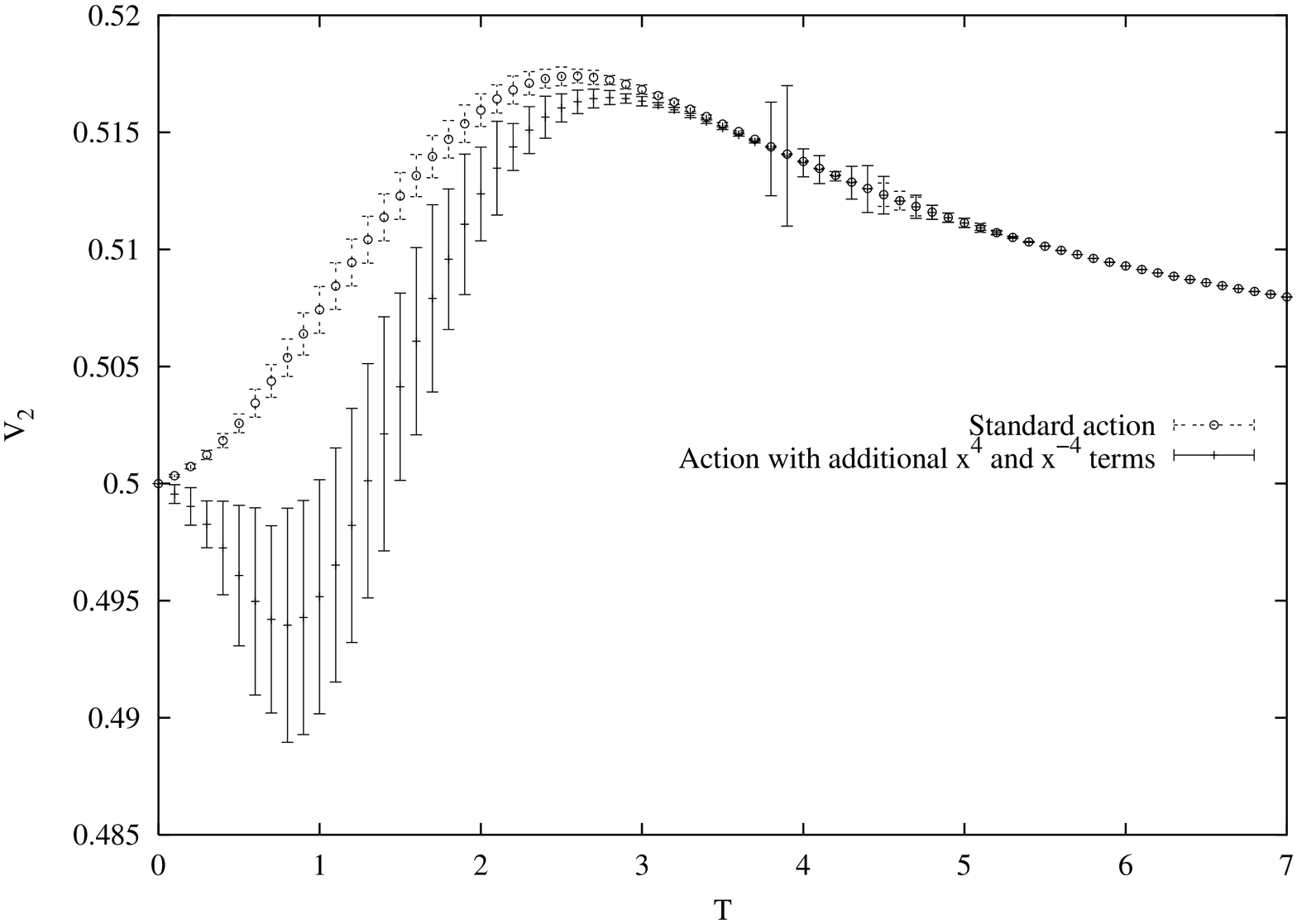}
\end{center}
\caption{Same as Fig.[\ref{Fig_e44_m}] for parameter $\tilde{v}_{2}$. }
\label{Fig_e44_v2}
\end{figure}

\begin{figure}[thb]
\vspace{9pt}
\begin{center}
\includegraphics[scale=0.4,angle=270]{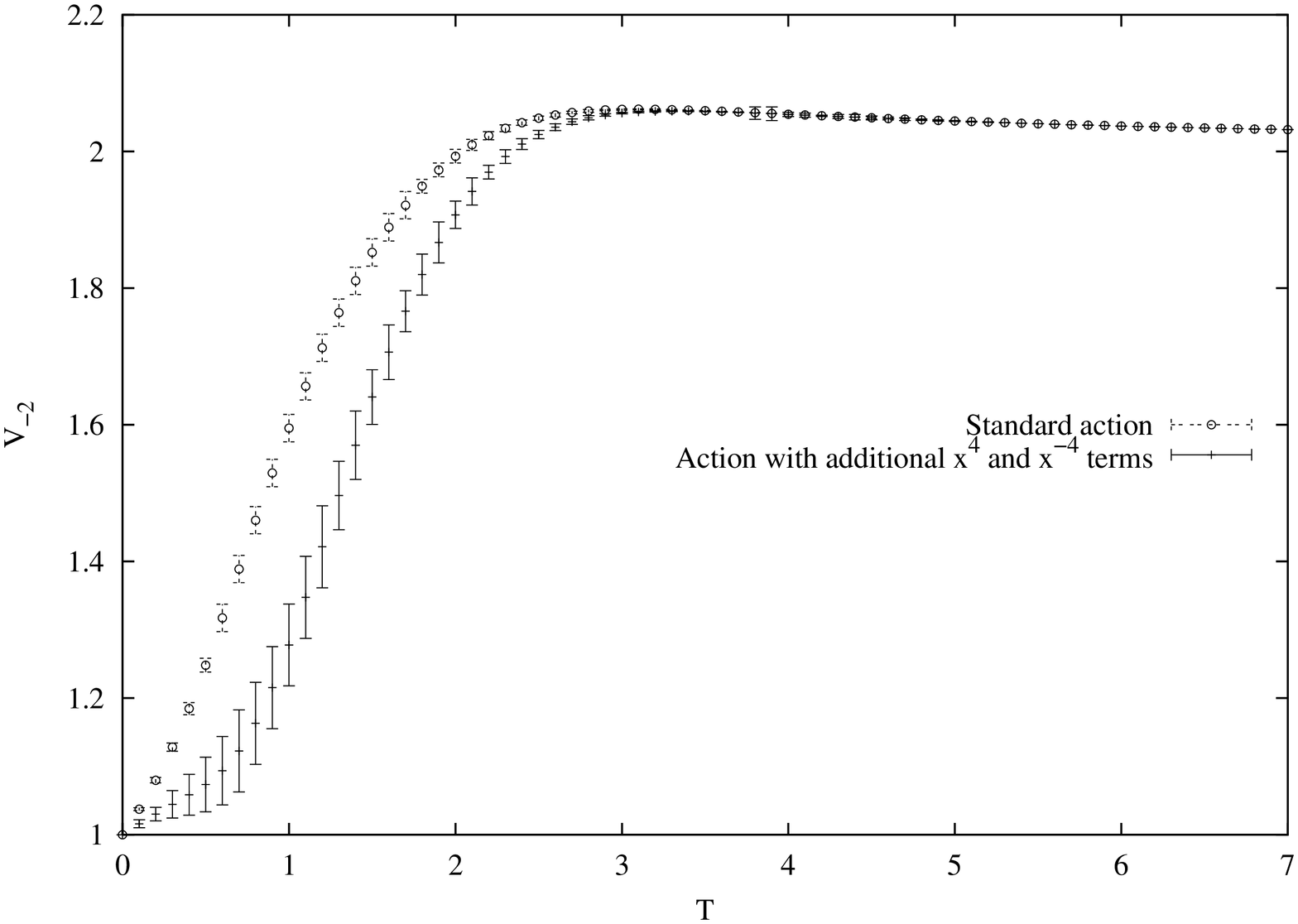}
\end{center}
\caption{Same as Fig.[\ref{Fig_e44_m}] for parameter $\tilde{v}_{-2}$. }
\label{Fig_e44_v-2}
\end{figure}

\begin{figure}[thb]
\vspace{9pt}
\begin{center}
\includegraphics[scale=0.4,angle=270]{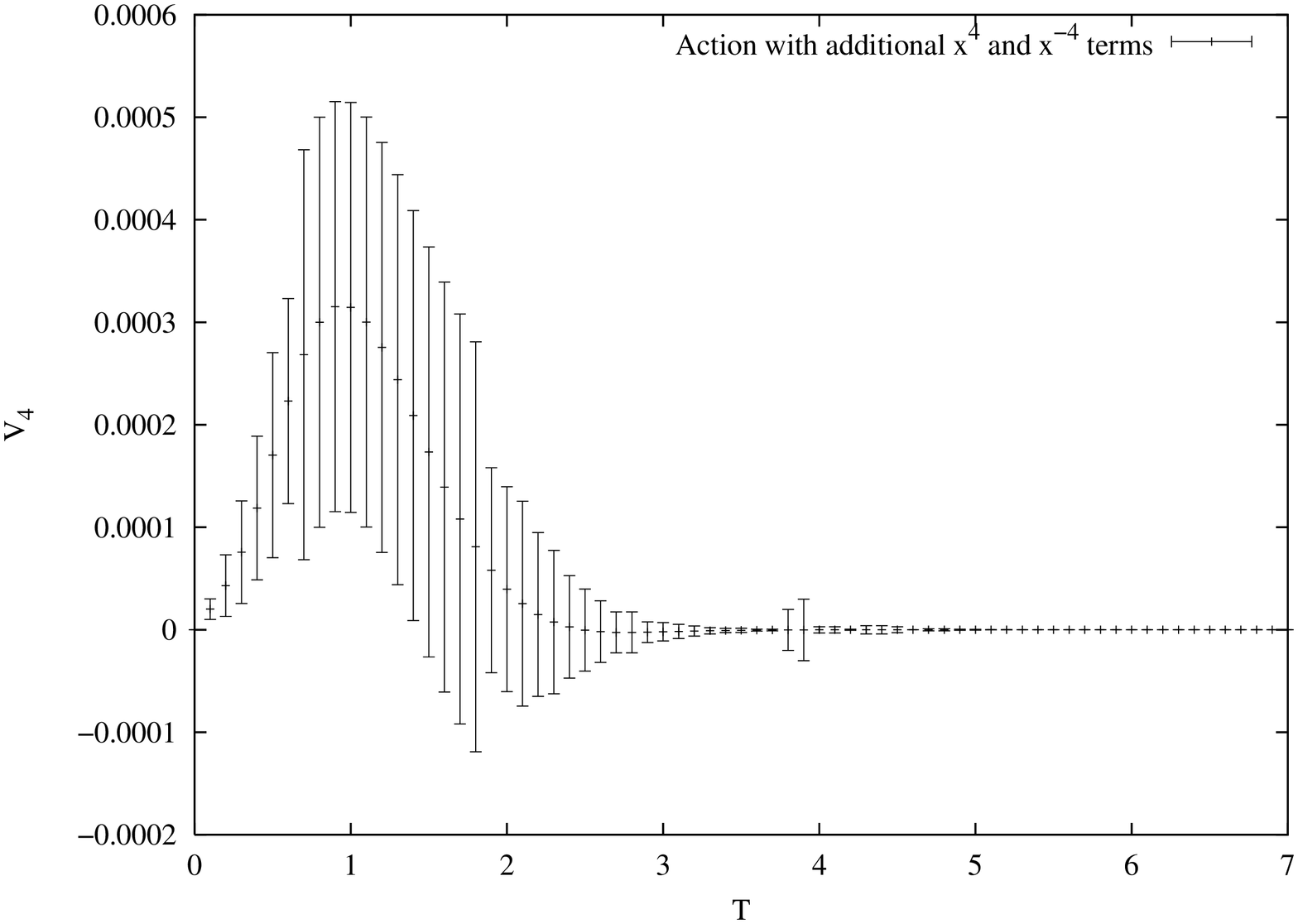}
\end{center}
\caption{Same as Fig.[\ref{Fig_e44_m}] for parameter $\tilde{v}_{4}$. }
\label{Fig_e44_v4}
\end{figure}

\begin{figure}[thb]
\vspace{9pt}
\begin{center}
\includegraphics[scale=0.4,angle=270]{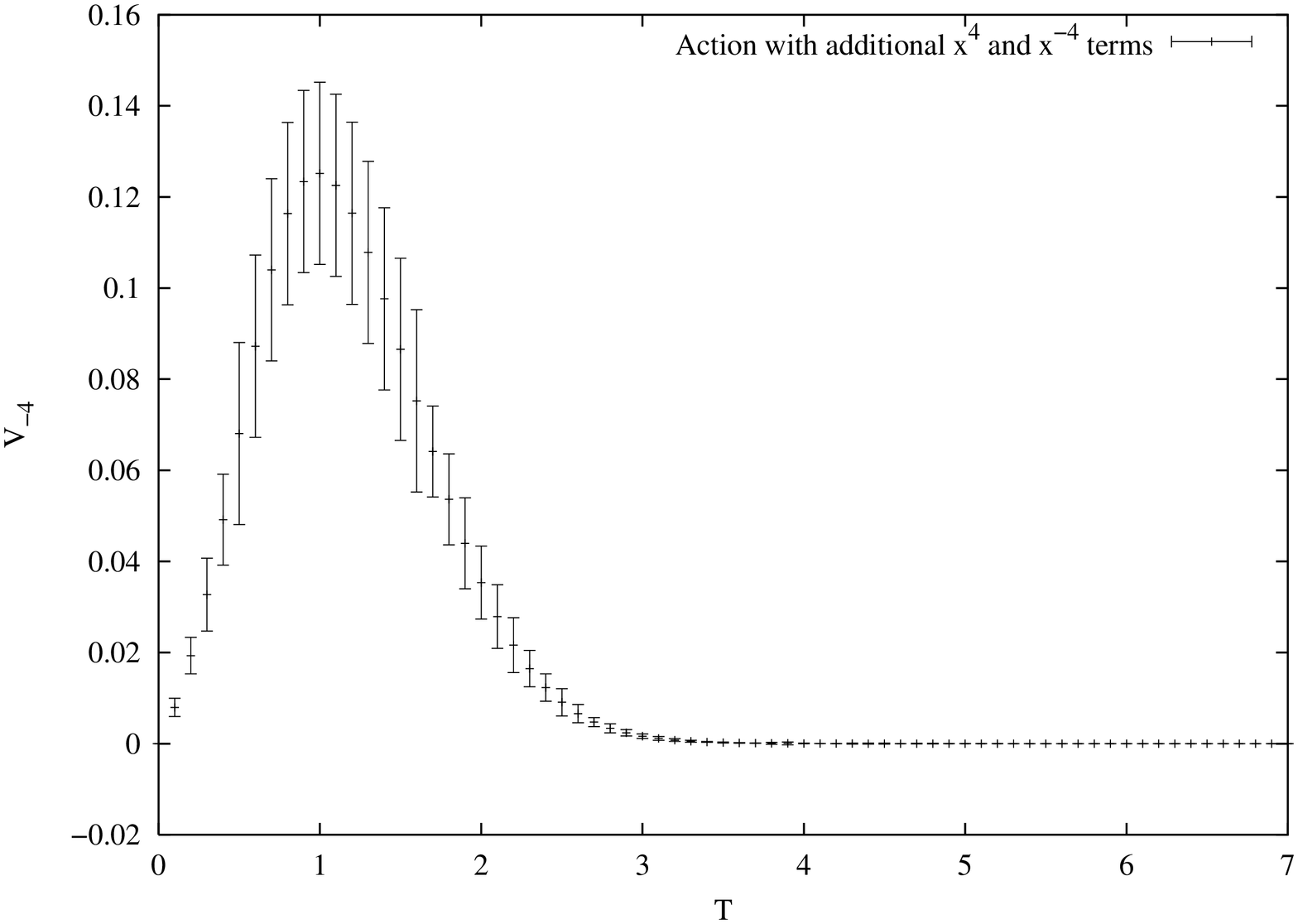}
\end{center}
\caption{Same as Fig.[\ref{Fig_e44_m}] for parameter $\tilde{v}_{-4}$. }
\label{Fig_e44_v-4}
\end{figure}

\begin{figure}[thb]
\vspace{9pt}
\begin{center}
\includegraphics[scale=0.4,angle=270]{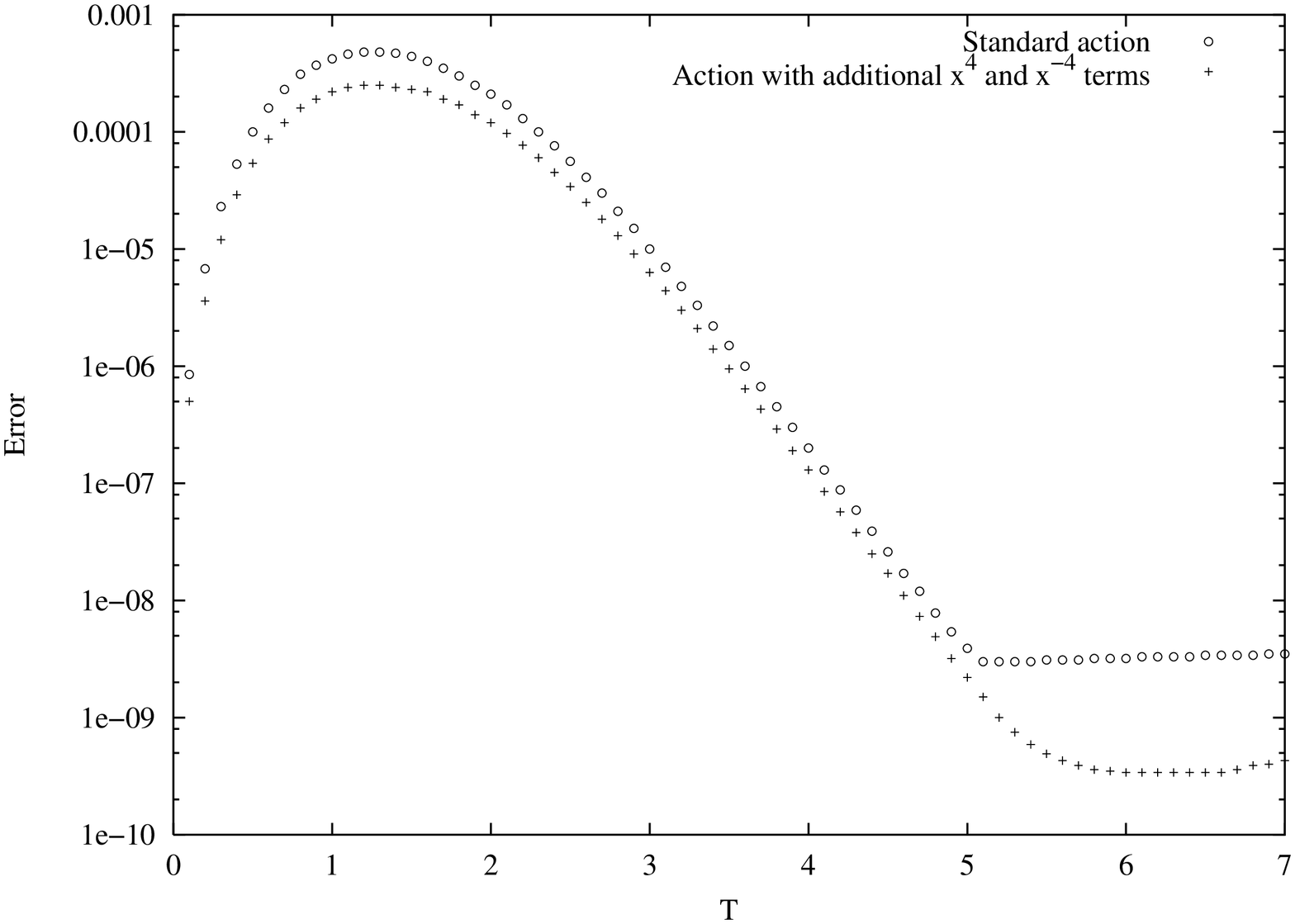}
\end{center}
\caption{Same as Fig.[\ref{Fig_e44_m}]. Global relative error 
of quantum action $\Sigma_{ij}$. }
\label{Fig_e44_S}
\end{figure}

\section{Use of the quantum action}
\label{sec:Use}

Like the standard effective action \cite{EffAct}, the optimized expansion of effective action \cite{OptimExp}, the Gaussian effective action \cite{GaussEffAct}, the Feynman-Kleinert effective action {\cite{FeynmanKlein} also the quantum action has been constructed for the purpose to study quantum phenomena, which have its origin in classical physics. Prototype examples are chaos and instantons. It is well known that classical chaos has no direct analogue in quantum physics. One reason is that in Q.M. one cannot define a point in phase space (due to Heisenberg's uncertainty relation), hence one 
cannot define Lyapunov exponents from diverging trajectories, 
thus quantum chaos 
cannot be represented quantitatively via Poincar\'e sections.
In Ref.\cite{JonaLasinio} the effective action has been used to describe 
classically chaotic quantum systems.
A comparison of Poincar\'e sections from the classical action and the quantum action has been presented in Ref.\cite{Q3}. 
A problem similar to that of a proper definition of quantum chaos exists also for the definition of instantons in quantum physics. Classical instantons are solutions going from one potential minimum to another. 
They start from the location of the potential minimum and with kinetic energy zero. Due to Heisenberg's uncertainty relation, there is no rigorous analogon in quantum physics. The physics of instantons and its relation to tunneling has been discussed in Refs.\cite{Coleman,Rajaraman}. For a review on the use of the standard effective action see  Ref.\cite{Leggett}.
A comparison of instantons from the classical action versus the quantum action has been given in Ref.\cite{Q2}.  
Because tunneling in Q.M. is very closely related to instantons, the quantum action is expected to shed new light into the phenomenon of 
Q.M. tunneling. Below we give a survey of topics in different areas of physics
where the use of the different kinds of effective action as well as the quantum action should be useful.

\begin{figure}[thb]
\vspace{9pt}
\begin{center}
\includegraphics[scale=0.4,angle=270]{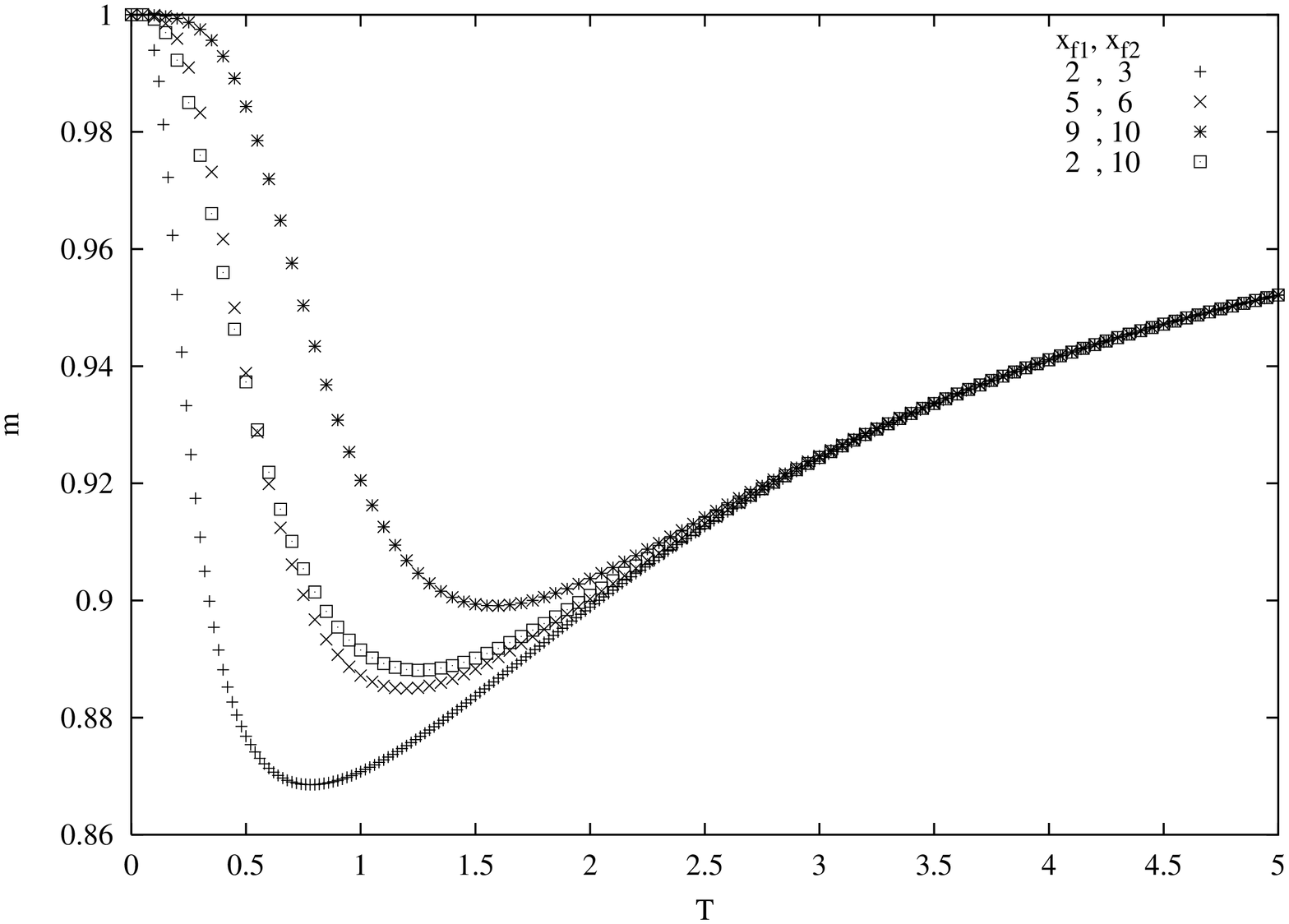}
\end{center}
\caption{Inverse square potential. Mass $\tilde{m}$ vs. transition time $T$. Dependence of parameters of quantum action on location of final boundary points: $x_{f}$ - 100 points in various intervals. 
Initial boundary point $x_{i} = 0.3$. }
\label{Fig_eevh_m}
\end{figure}

\begin{figure}[thb]
\vspace{9pt}
\begin{center}
\includegraphics[scale=0.4,angle=270]{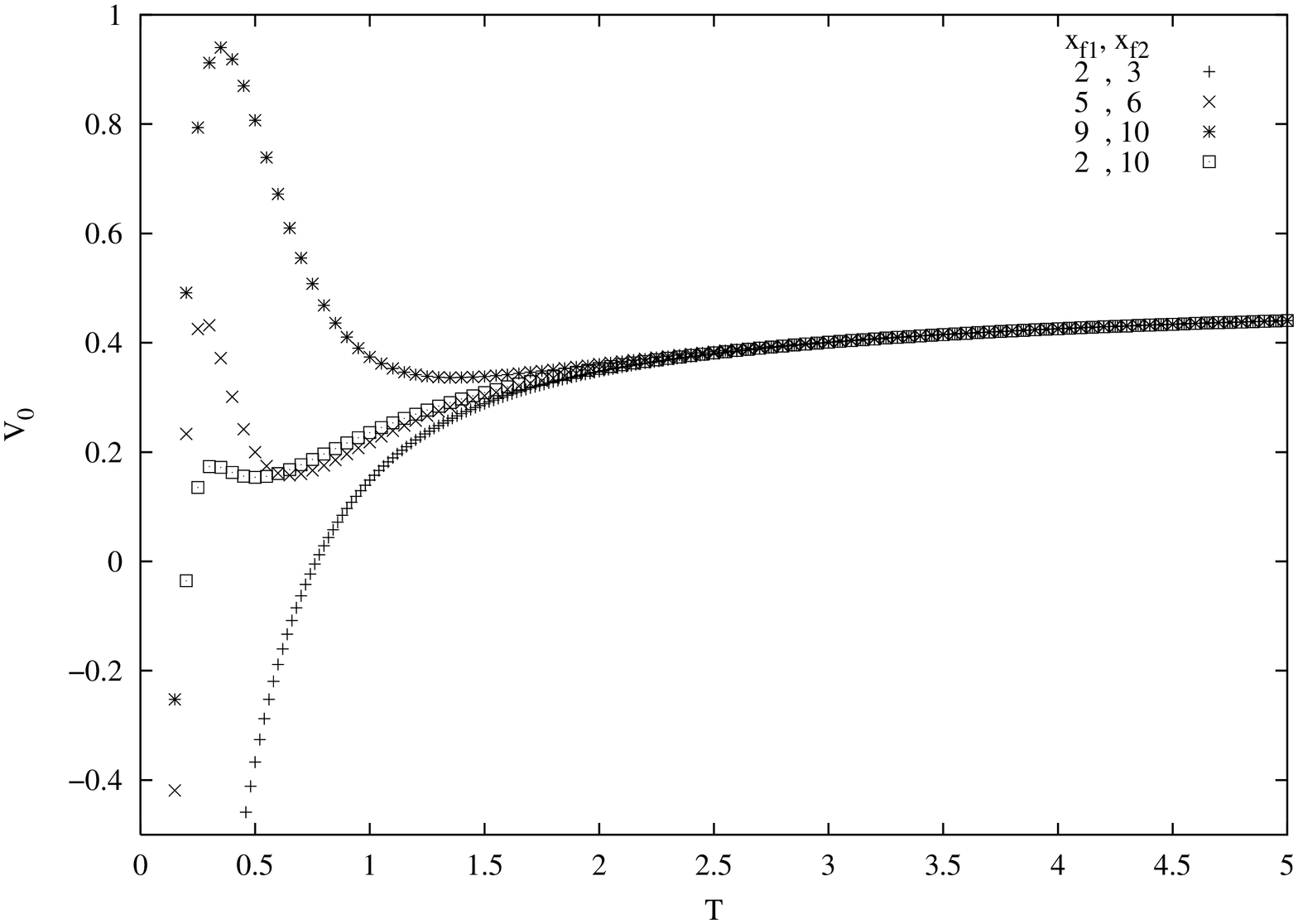}
\end{center}
\caption{Same as Fig.[\ref{Fig_eevh_m}] for parameter $\tilde{v}_{0}$. }
\label{Fig_eevh_v0}
\end{figure}

\begin{figure}[thb]
\vspace{9pt}
\begin{center}
\includegraphics[scale=0.4,angle=270]{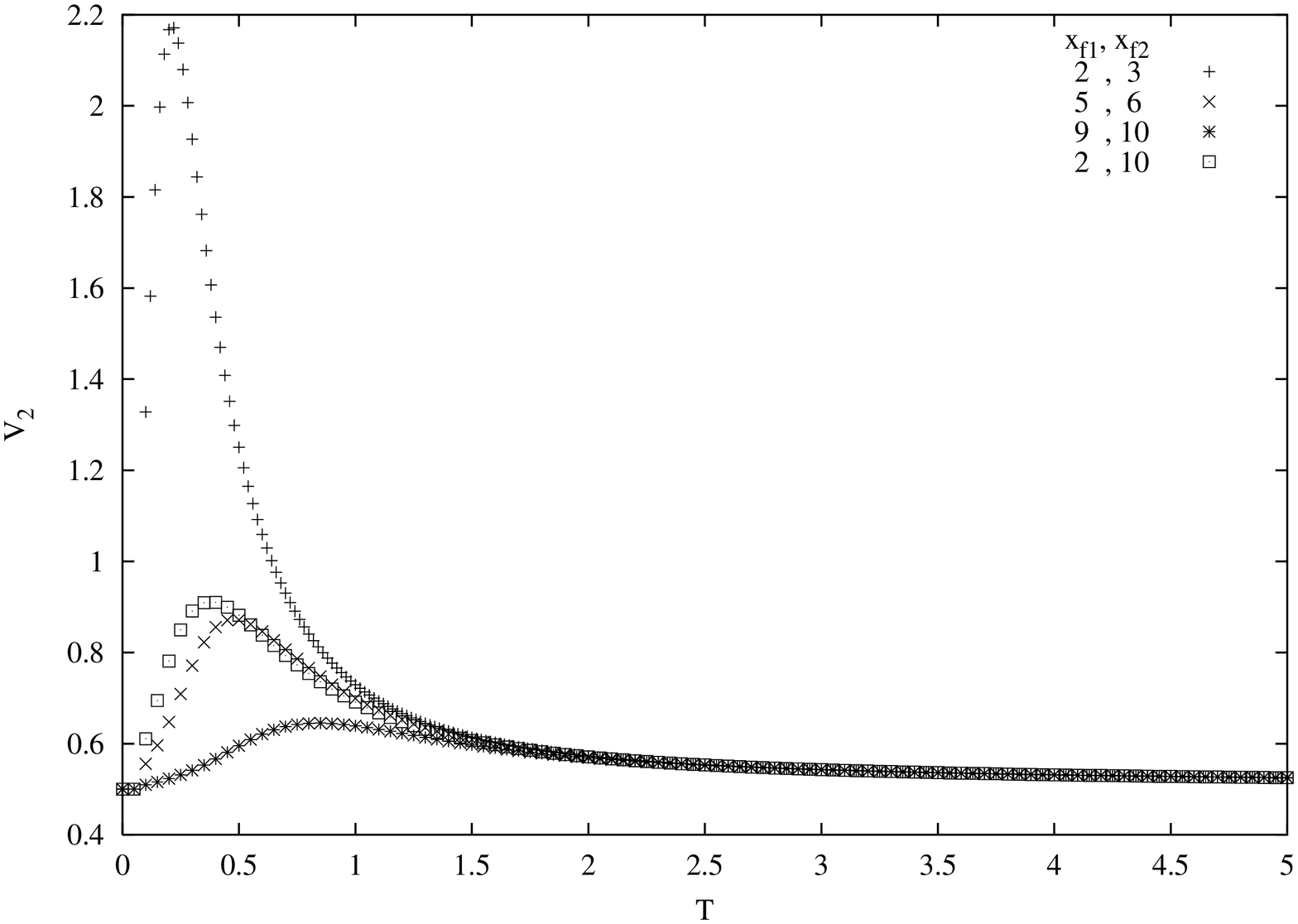}
\end{center}
\caption{Same as Fig.[\ref{Fig_eevh_m}] for parameter $\tilde{v}_{2}$. }
\label{Fig_eevh_v2}
\end{figure}

\begin{figure}[thb]
\vspace{9pt}
\begin{center}
\includegraphics[scale=0.4,angle=270]{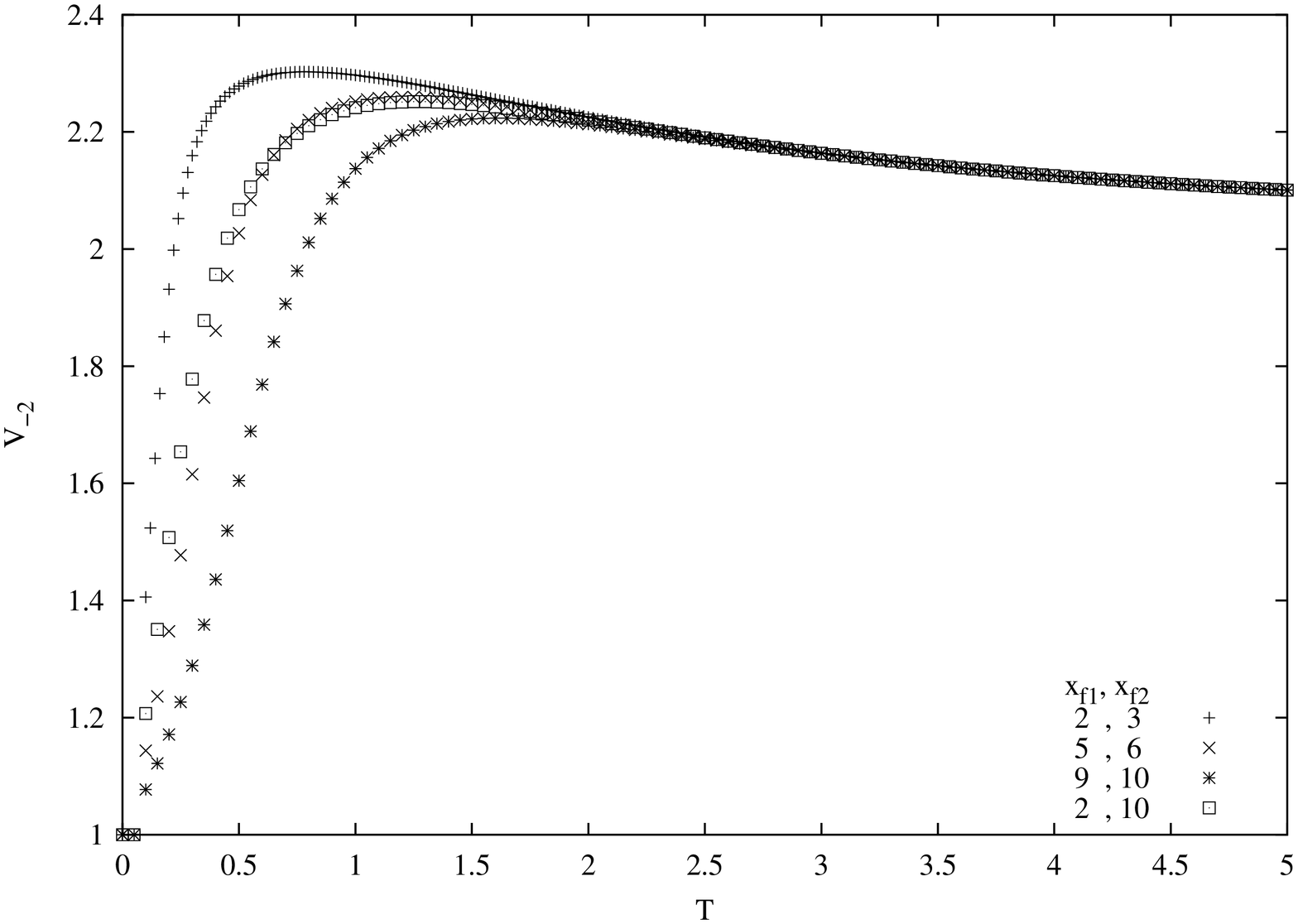}
\end{center}
\caption{Same as Fig.[\ref{Fig_eevh_m}] for parameter $\tilde{v}_{-2}$. }
\label{Fig_eevh_v-2}
\end{figure}

\subsection{High energy physics}
\label{sec:HighEner} 

(i) Renormalisation. The quantum action may turn out to provide a new definition of renormalisation. In conventional terminology, renormalisation means to extract physical observable parameters like mass, coupling etc. obtained from an interacting quantum field theory at the continuum limit.
Those physical parameters differ from the so-called bare parameters and this difference represents the effects of the interaction. 
Very similar to this, the quantum action has parameters (mass, potential parameters) different from the classical action and this difference 
represents the effects of quantum mechanics: Q.M. fluctuations occuring in the path integral are represented by a single path, however, for a particle with different properties of mass and interaction. 
In order to explore the use of the quantum action as means of renormalisation,
one could construct the quantum action for a many-body system (like a chain of coupled anharmonic oscillators) and explore its quantum mechanical continuum limit. In quantum field theory, one carries out renormalisation by computing n-point vertex functions, which represent vacuum-to-vacuum transition amplitudes for transition time $T \to \infty$.
Exactly in the limit $T \to \infty$, the existence of the quantum action has been proven rigorously in Q.M., although so far only in the case of a single particle system.    

(ii) Cosmology and inflationary scenario of early universe. 
Inflation involves potentials with several minima and instanton solutions. 
The instanton starts out as a quantum instanton and eventually turns into a classical instanton. This has effects on the subsequent formation of galaxies \cite{Staro79,Khlo98,Kolb91}. Using the effective potential or the quantum potential (potential of the quantum action) in general creates a potential different from the classical one. In particular, it may have minima being absent in the classical potential. Consequently, this may create instantons being quite different from the classical instanton (in Ref.\cite{Q2} instantons from the quantum action were found to be "softer" than the corresponding classical instanton). Such quantum effects of the instanton may influence the outcome of the galaxy formation at the end of inflation.

(iii) Hot and dense nuclear matter. Instantons are believed to play an important role in hot nuclear matter in the quark-gluon plasma phase \cite{Shur88}. 
Instantons are important also for the mechanism of chiral symmetry breaking and for t'Hoofts solution of the U(1) problem. 
 
(iv) Neutrino oscillations. The process of oscillations of neutrino flavors 
\cite{NeutrOsc1,NeutrOsc2,NeutrOsc3} may have to do with a process of tunneling in a potential with degenerate minima.

\subsection{Condensed matter physics} 
\label{sec:CondMatt}

(i) Quantum dots, semi-conductor and quantum chaos. 
Advancing the speed of microprocessors may have technological obstacles but also physical limits. When reducing the size of a chip one soon may enter the regime where quantum laws rule. Quantum chaos may become a very important issue, because it can hamper the flow of electric currents. The quantitative determination of quantum chaos effects will possibly be of great importance for the development of future microprocessors. 
By use of some kind of effective action or the quantum action, one can study  quantum corrals formed by atoms and quantum dots in semi-conductors.  
In particular, this allows to study the temperature dependence of electron dynamics in atomic corrals, as well as for electrons moving in simple conductor-semiconductor-isolator geometries and to search for the possible presence of quantum chaos.

(ii) Josephson junctions and superconducting quantum interference. Superconducting quantum interference devices (SQUID)
have been used to demonstrate experimentally the phenomenon of quantum superposition in macroscopic states \cite{Friedman}. This involves Josephson junctions. The SQUID potential has a double-well structure. 
The effective potential and the quantum potential should be useful tools to analyze quantum superposition in terms of such potential involving quantum effects.

(iii) Quantum computers based on superconductors.
The symmetry of the order parameter in some triplet superconductors corrersponds to doubly degenerate chiral states.
Gulian et al.\cite{Gulian1,Gulian2} predict that this degeneracy can be lifted via macroscopic quantum tunneling. Instanton-like quantum behavior may become important. Triplet superconducters may be used as basic elements of quantum computers. Again the effective potential and the quantum potential should help to study quantum instantons and tunneling in such materials.

\subsection{Atomic physics} 
\label{sec:Atom}

(i) Analogy of classical chaos in quantum physics. Attempts have been made to use the effective action in order to characterize chaotic behavior in quantum systems \cite{JonaLasinio}. In a a similar way, the quantum action has been used also \cite{Q3}. This allows to construct of a phase space portrait and Poincar\'e sections for a quantum system in analogy to classical physics. From this one can obtain Lyapunov exponents and KAM surfaces for the quantum system. A potentially quite interesting system to explore is the Paul trap or similar traps.

(ii) Ultracold atoms in a billard formed by lasers. Trace formulas (Gutzwiller \cite{Gutz90} and generalisations ) have been used successfully \cite{Miln01,Fried01,Dembr01} to establish a relation between level densities and periodic semi-classical orbits. It would be interesting to compare predictions of trace formulas in the semi-classical regime with the predictions obtained from the effective action or the quantum action. 

(iii) Dynamical tunneling. Steck et al.\cite{Raizen} 
and Hensinger et al.\cite{Hensinger}  
have demonstrated experimentally the phenomenon of dynamical tunneling (where the classical transition is forbidden due to some conserved quantity different from energy). It has been realized by arrays of cold atoms. It has been observed that the presence of quantum chaos enhances the dynamical tunneling transition.
It would be instructive to reexamine dynamical tunneling using the phase space portrait constructed from a time-dependent effective action or quantum action.

\subsection{Chemistry} 
\label{sec:Chem}

Binding of macromolecules. In the process of chemical binding of macromolecules, often a double well potential plays a role. 
The effective action as well as the quantum action should be useful to find pathways in the formation of such macromolecules.

\section{Quantum action vs. effective action} 
\label{sec:QActEffAct}

What are the similarities of the standard effective action and the quantum action? Where are the differerences?

(i) The effective action is computed via loop expansion. This leads to non-localities. The effective action is given by an infinite series of terms. Those non-localities are manifested by higher order terms of time-derivatives 
[ $(\frac{dx}{dt})^{n}$, $(\frac{d^{2}x}{dt^{2}})^{m}$, etc.] 
The series of such higher terms is infinite.
This causes problems because the solution of the Euler-Lagrange equation of motion requires infinitely many boundary conditions. 
Moreover, such series does not converge, it is only asymptiotically valid.  
To make practical use of the effective action one proceeds by retaining a few low order terms. However, for the purpose to study quantum chaos \cite{JonaLasinio}, this may be problematic, because in 
chaotic dynamics "small terms" may cause large effects. 
None of such non-localities occur in the quantum action.

\begin{figure}[thb]
\vspace{9pt}
\begin{center}
\includegraphics[scale=0.4,angle=270]{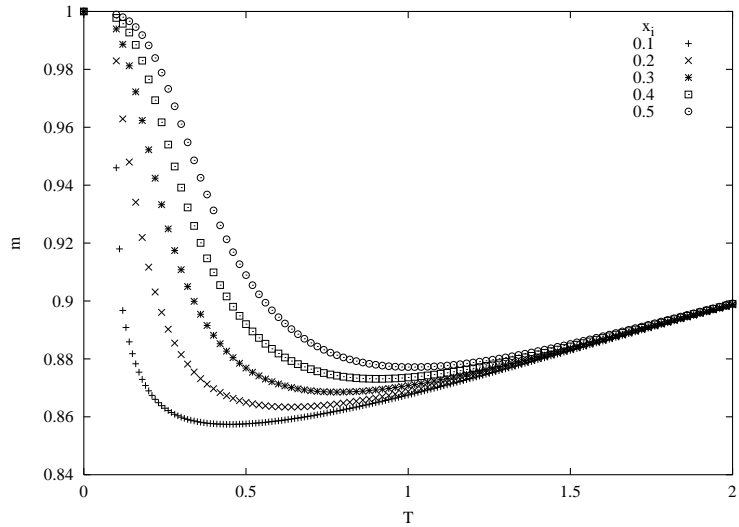}
\end{center}
\caption{Inverse square potential. Mass $\tilde{m}$ vs. transition time $T$. Dependence of parameters of quantum action on location of initial boundary point $x_{i}$. Final boundary points $x_{f}$ - 100 points in interval $[2,3]$. }
\label{Fig_eevh23_m}
\end{figure}

\begin{figure}[thb]
\vspace{9pt}
\begin{center}
\includegraphics[scale=0.4,angle=270]{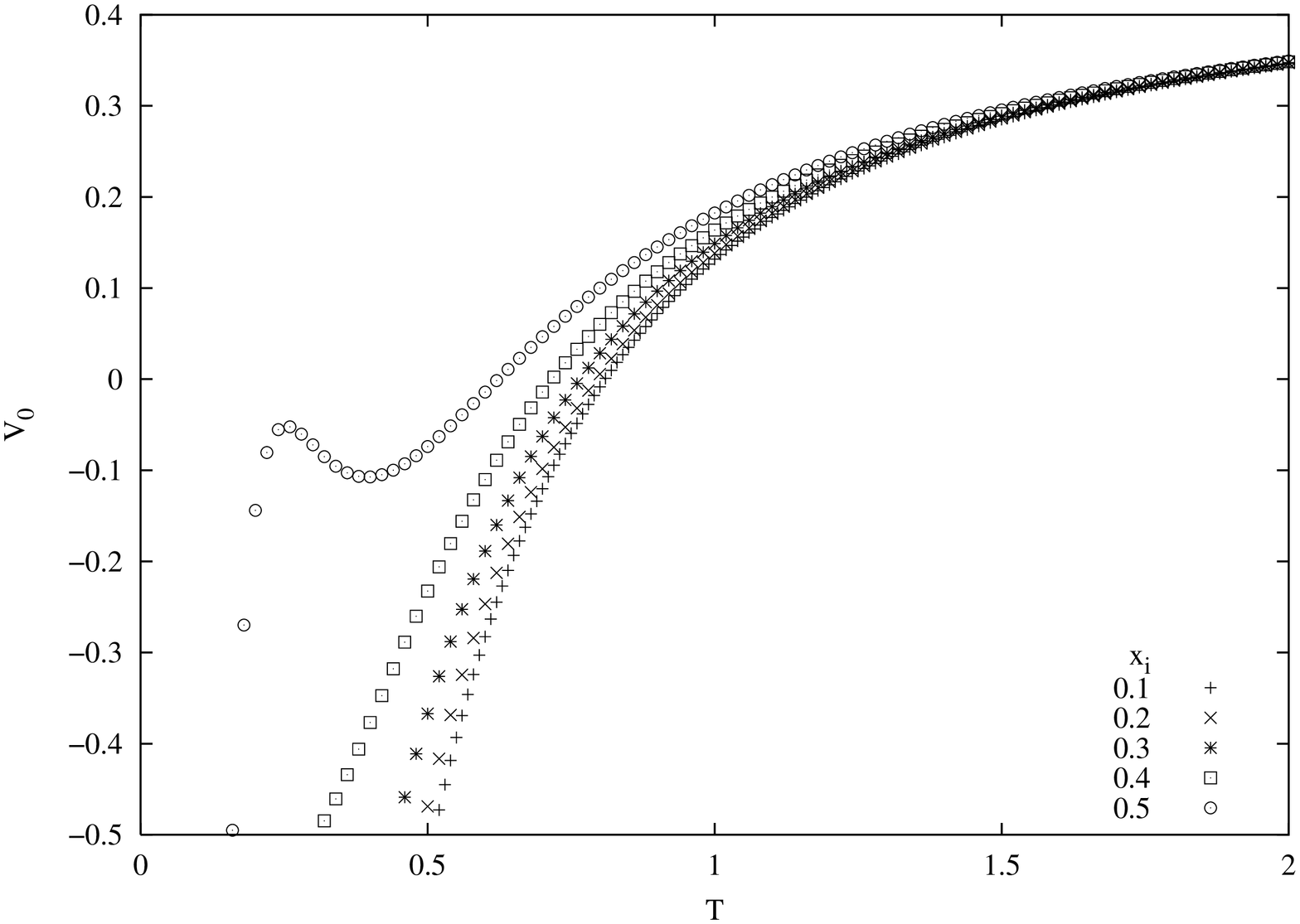}
\end{center}
\caption{Same as Fig.[\ref{Fig_eevh23_m}] for $\tilde{v}_{0}$. }
\label{Fig_eevh23_v0}
\end{figure}

\begin{figure}[thb]
\vspace{9pt}
\begin{center}
\includegraphics[scale=0.4,angle=270]{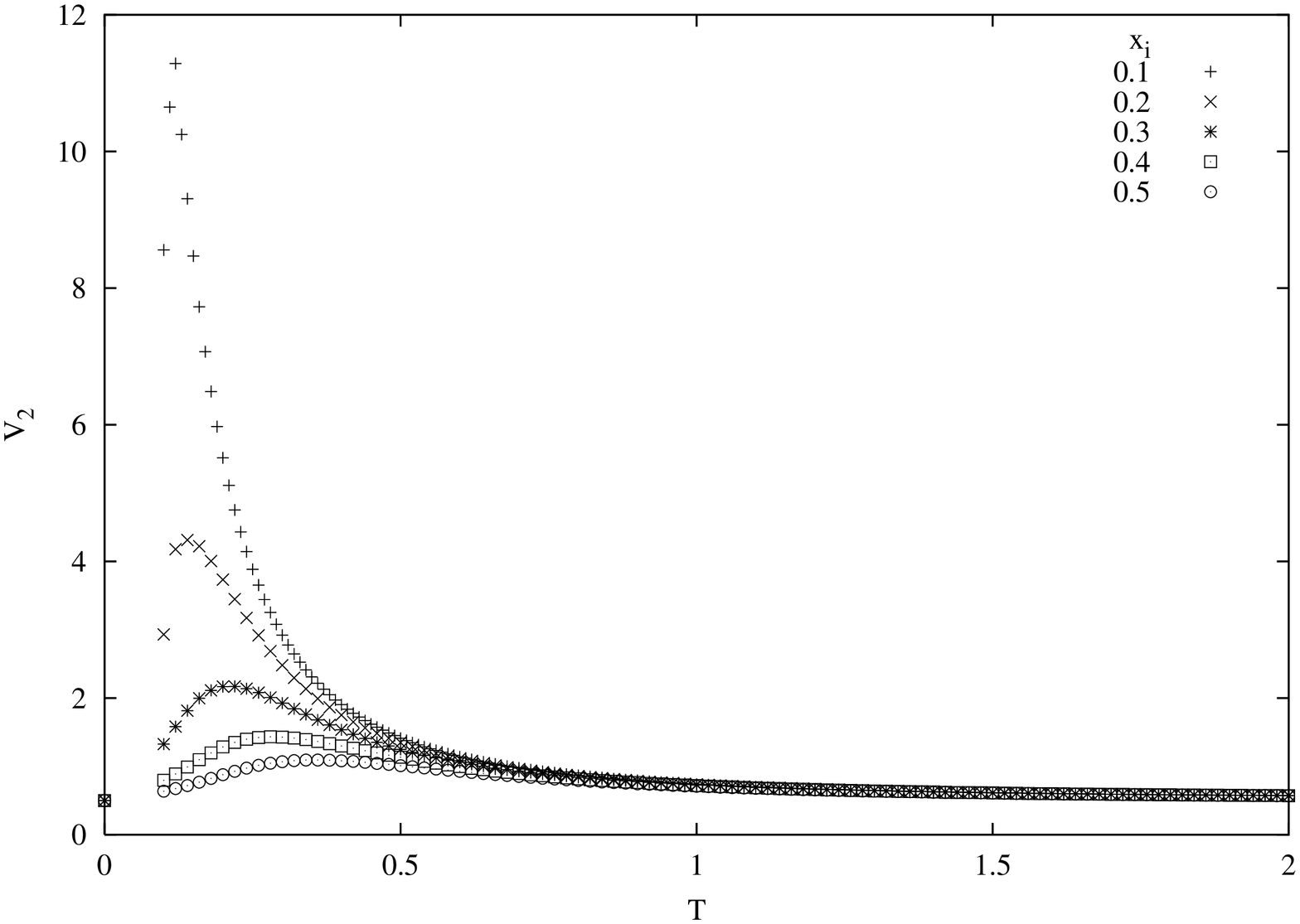}
\end{center}
\caption{Same as Fig.[\ref{Fig_eevh23_m}] for $\tilde{v}_{2}$. }
\label{Fig_eevh23_v2}
\end{figure}

\begin{figure}[thb]
\vspace{9pt}
\begin{center}
\includegraphics[scale=0.4,angle=270]{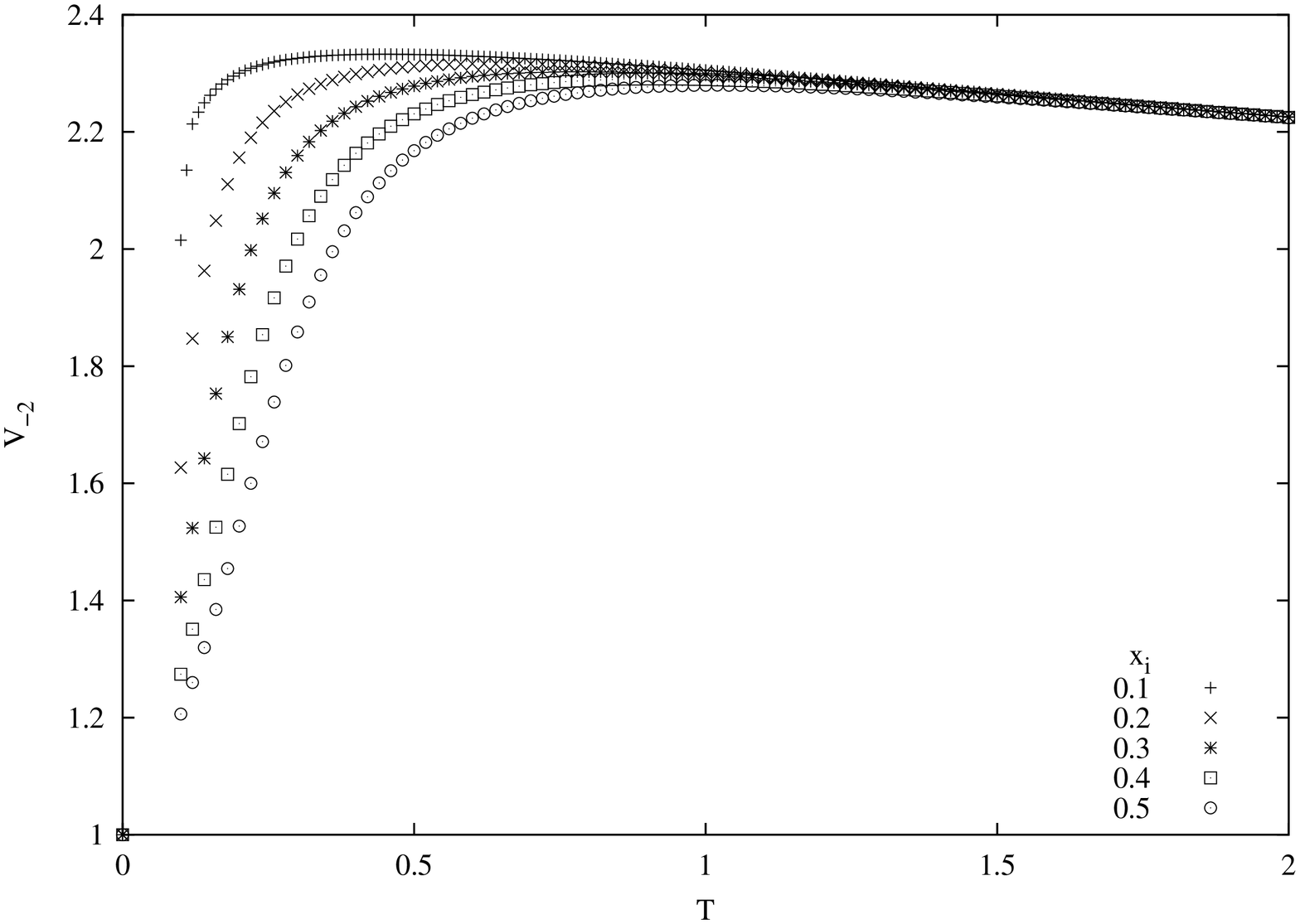}
\end{center}
\caption{Same as Fig.[\ref{Fig_eevh23_m}] for $\tilde{v}_{-2}$. }
\label{Fig_eevh23_v-2}
\end{figure}

(ii) While the standard effective action does not reproduce a potential of double well shape, the Gaussian effective action produces an effective potential with double well shape \cite{GaussEffAct,OptimExp}. This is also the case for the Feynman-Kleinert type of effective action \cite{FeynmanKlein}.  

(iii) Q.M. tunneling has no classical counterpart. It is closely related to instantons. One should note that the particular tunneling process and also the instantons depend on the shape of the potential, its barrier height as well as the position of minima. Because different types of effective action differ in their corresponding effective potential, consequently they differ in their 
tunneling amplitudes. 

(iv) The quantum action shares with the effective action the property that the ground state energy is given by the minimum of the quantum, respectively effective potential. But the quantum action is distinguished by the property of coincidence of location(s) of maximum(a) of the ground state wave function and 
minimum(a) of the quantum potential \cite{Q5}.

\section{Validity of the quantum action}
\label{sec:ValQAct}

The quantum action has been originally proposed as a conjecture. 
A priori, it is not evident that such quantum action exists.
By now the existence of the quantum action has been established
in the following cases: 

(i) Harmonic oscillator. In this case the quantum action is identical to the classical action. 

(ii) In the limit when the transition time $T \to 0$, the quantum action exists. Dirac has noticed that the path integral in this limit is dominated by the contribution from the classical trajectory. Hence the quantum action coincides with the classical action.

(iii) In imaginary time (necessary to describe thermodynamics at finite temperature) and going to the limit $T \to \infty$, the quantum action exists. In general it is quite different from the classical action.

This leads to the question: What about finite transition time? 
Does the quantum action exist for arbitray finite transition time? Does it parametrize Q.M. transition amplitudes for all $x_{in}$ to $x_{fi}$ equally well? In general, the quantum action has to be determined non-perturbatively by numerical computations. What can be said about about stability, convergence and errors of such procedure? Those questions are the subject of the paper. We have chosen to study this at hand of the inverse square potential in one dimension.

\begin{figure}[thb]
\vspace{9pt}
\begin{center}
\includegraphics[scale=0.4,angle=270]{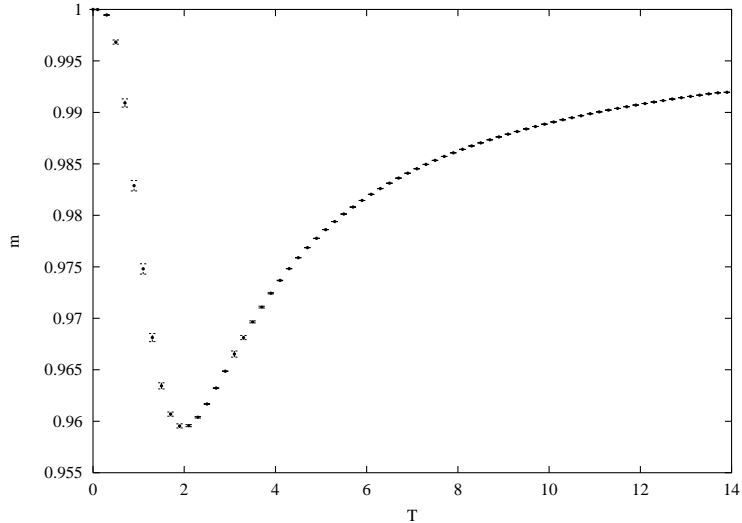}
\end{center}
\caption{Inverse square potential. Mass $\tilde{m}$ vs. transition time $T$. Dependence of parameters of the quantum action on boundary points: $x_{i}$ - 10 points in interval $[1.5, 2.5]$; $x_{f}$ - 10 points in interval $[1.1, 2.1]$. }
\label{Fig_estand_m}
\end{figure}

\begin{figure}[thb]
\vspace{9pt}
\begin{center}
\includegraphics[scale=0.4,angle=270]{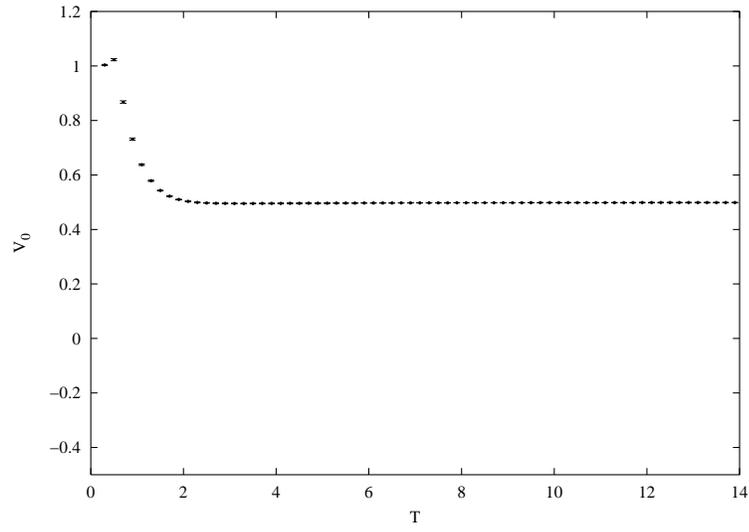}
\end{center}
\caption{Same as Fig.[\ref{Fig_estand_m}] for $\tilde{v}_{0}$. 
One observes $v_0 \to 0.5$, consistent with analytic result (minimum 
of quantum potential goes to $E_{gr}$). }
\label{Fig_estand_v0}
\end{figure}

\begin{figure}[thb]
\vspace{9pt}
\begin{center}
\includegraphics[scale=0.4,angle=270]{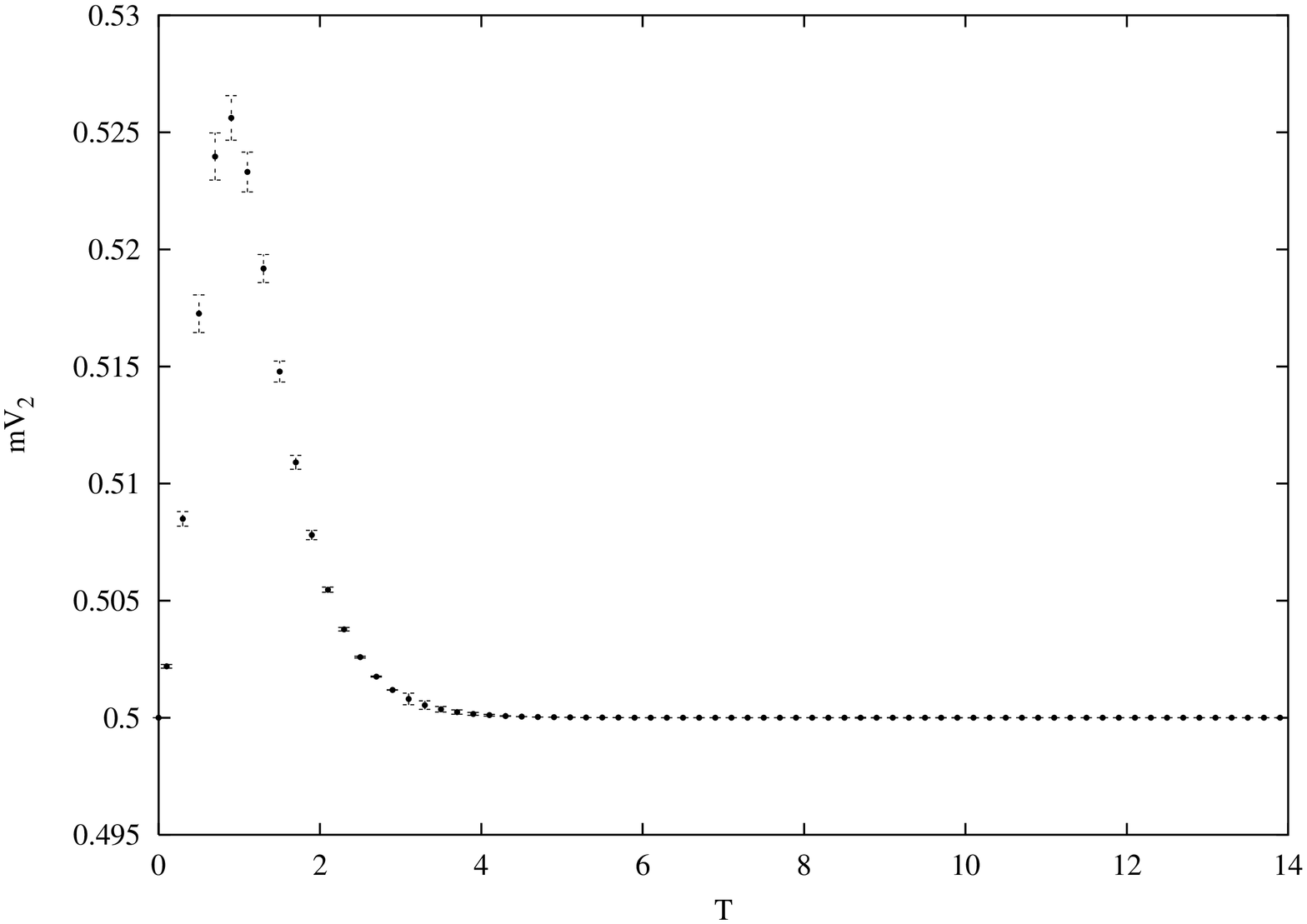}
\end{center}
\caption{Same as Fig.[\ref{Fig_estand_m}] for $\tilde{m} \tilde{v}_{2}$. 
One observes $\tilde{m} \tilde{v}_{2} \to 0.5$, 
consistent with analytic result. }
\label{Fig_estand_mv2}
\end{figure}

\begin{figure}[thb]
\vspace{9pt}
\begin{center}
\includegraphics[scale=0.4,angle=270]{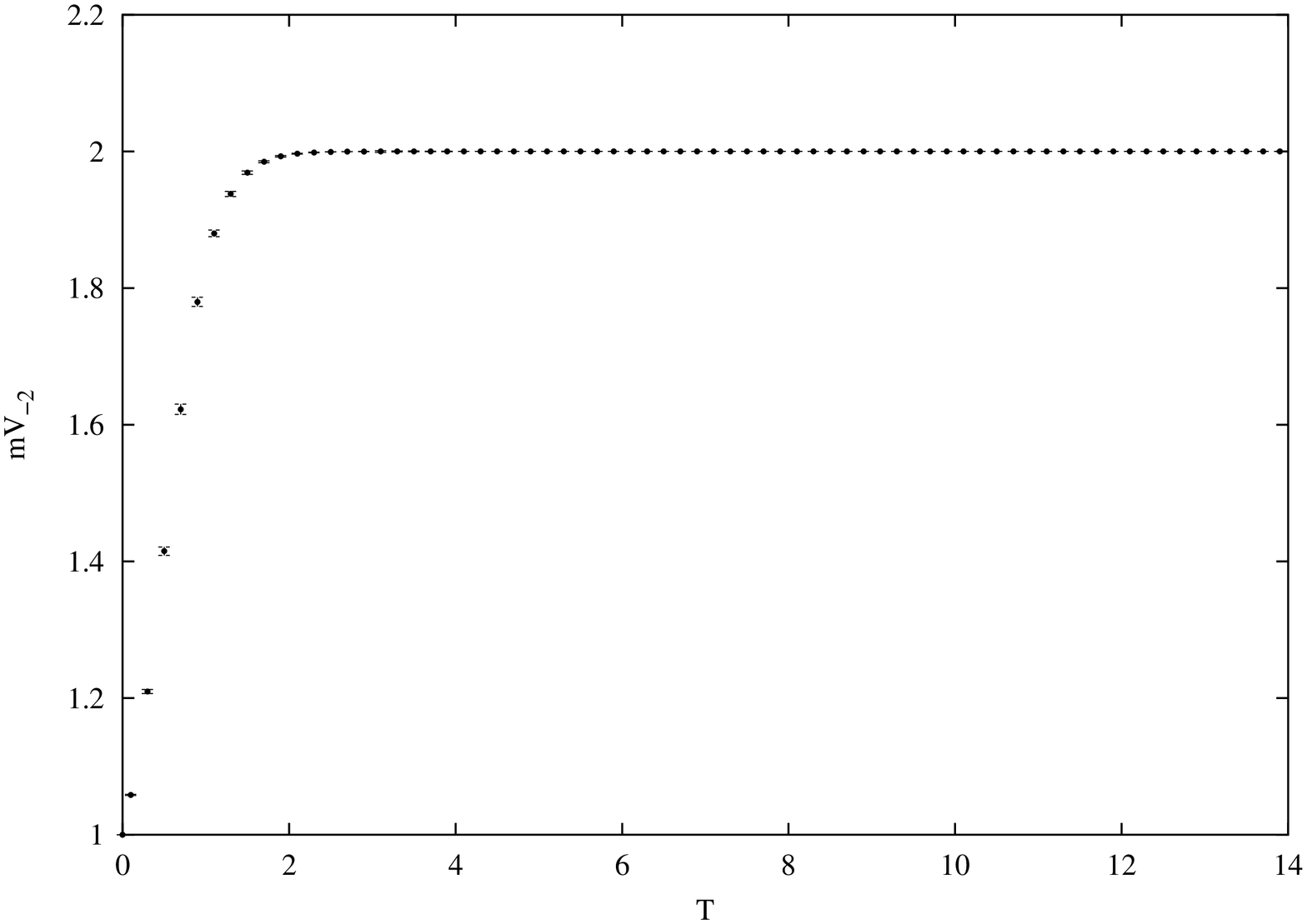}
\end{center}
\caption{Same as Fig.[\ref{Fig_estand_m}] for $\tilde{m} \tilde{v}_{-2}$. 
One observes that $\tilde{m} \tilde{v}_{-2} \to 2$, consistant with analytic result. }
\label{Fig_estand_mv-2}
\end{figure}

\begin{figure}[thb]
\vspace{9pt}
\begin{center}
\includegraphics[scale=0.4,angle=270]{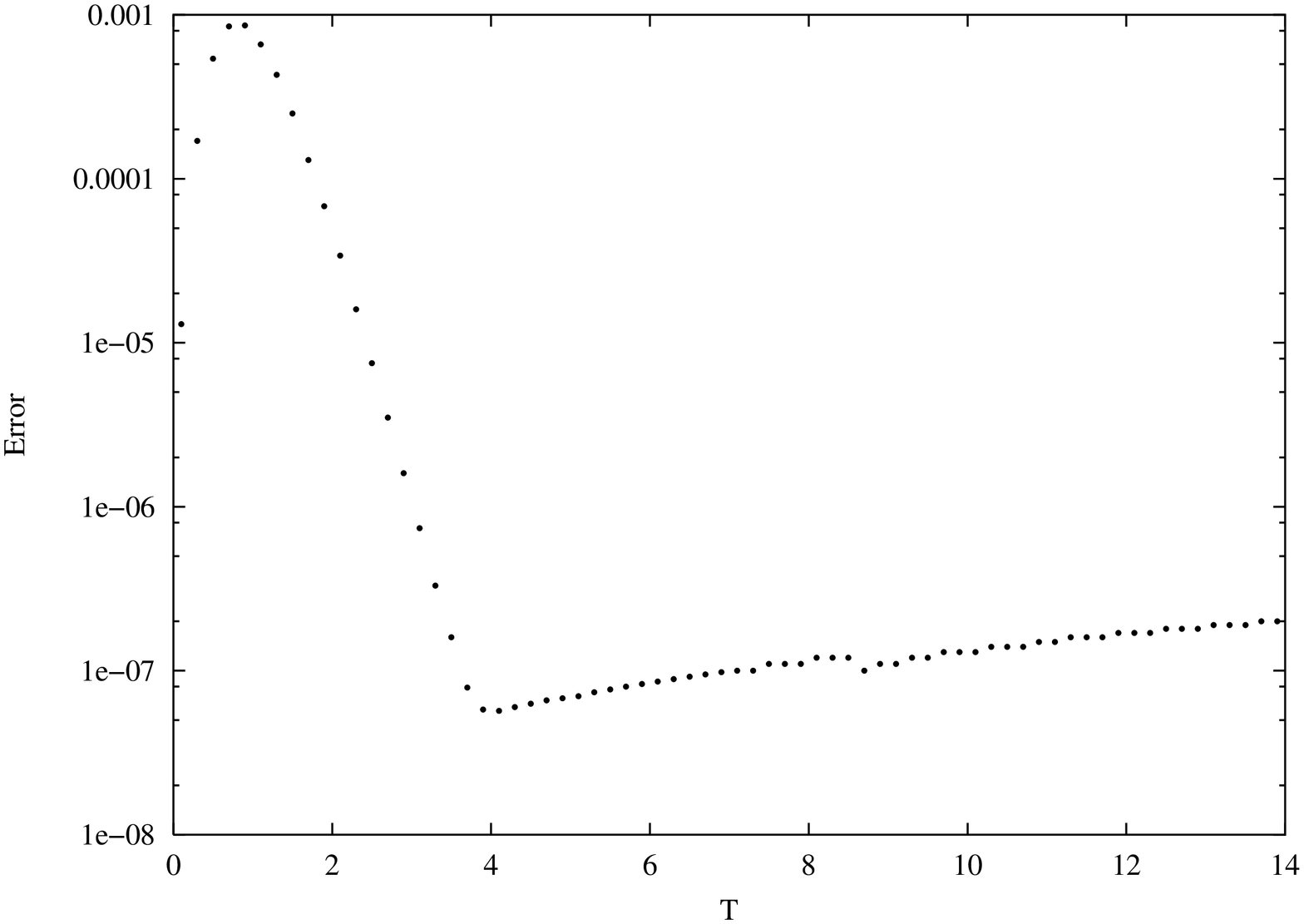}
\end{center}
\caption{Same as Fig.[\ref{Fig_estand_m}]. Global relative error of $G_{ij}$. }
\label{Fig_estand_G}
\end{figure}

\section{Inverse square potential}
\label{sec:InvSqPot}

We consider the following classical potential in 1-D
\be
\label{ClassPot}
V(x) = \frac{1}{2} m \omega^{2} ~ x^{2} + g ~ x^{-2} 
= v_{2} ~ x^{2} + v_{-2} ~ x^{-2} ~ .
\end{equation}
The corresponding classical action is given by
$S = \int dt ~ \frac{1}{2} m \dot{x}^{2} - V(x)$, while the 
classical action in imaginary time ($t \to -it$) is given by 
$S_{E} = \int dt ~ \frac{1}{2} m \dot{x}^{2} + V(x)$ 
(note: following conventions used in physics, the overall minus sign has been dropped. It reappears in the weight factor $\exp[-S_{E}]$). 
The potential is parity symmetric.
Because it has an infinite barrier (for $g > 0$) at the origin,
the system at $x < 0$ is separated from the system at $x > 0$. We consider only the motion in the domain $x > 0$. The potential is shown in Fig.[\ref{Fig_PotWave}]. We have chosen to consider the inverse square potential, because of the distinct feature that the corresponding quantum mechanical transition amplitudes are known analytically \cite{Khandekar,Schul.32}.
They are given by
\be
G(b,T;a,0)=\frac{m\omega \sqrt{ab} }{ i \hbar \sin(\omega T) } 
\exp \left\{ \frac{i m \omega} {2 \hbar} (b^{2} + a^{2}) \cot(\omega T) \right\}
I_{\gamma} \left( \frac{m \omega a b} {i \hbar \sin(\omega T) } \right) ~ ,
\end{equation}
where $I_{\gamma}$ is the modified Bessel function and 
\be
\gamma = \frac{1}{2} \left( 1 + \frac{8 m g}{\hbar^{2} } \right)^{1/2} ~ .
\end{equation}
The numerical studies discussed below have been done in imaginary time. The Euclidean transition amplitude reads
\be
G_{E}(b,T;a,0)=\frac{m\omega \sqrt{ab} }{ \hbar \sinh(\omega T) } 
\exp \left\{- \frac{ m \omega} {2 \hbar} (b^{2} + a^{2}) \coth(\omega T) \right\}
I_{\gamma} \left( \frac{m \omega a b} {\hbar \sinh(\omega T) } \right) ~ .
\end{equation}
The transition amplitude contains all information on the spectrum and wave functions. For example, by going to the limit $T \to \infty$ (Feynman-Kac limit), the Euclidean transition amplitude is projected onto the ground state.
One finds
\begin{eqnarray}
G_{E}(b,T;a,0) \longrightarrow_{T \to \infty} &&
Z_{0} ~ \exp[-\omega (1+\gamma)T ] 
\nonumber \\
&& b^{1/2+\gamma} ~ \exp[- \frac{m \omega}{2 \hbar} b^{2} ] 
\nonumber \\
&& a^{1/2+\gamma} ~ \exp[- \frac{m \omega}{2 \hbar} a^{2} ] ~ ,
\end{eqnarray}
where
\be
Z_{0} = \frac{2 m \omega}{\hbar \Gamma(\gamma+1) } 
\left( \frac{m \omega}{\hbar} \right)^{\gamma} ~ .
\end{equation}
One reads off the ground state energy and wave function
\begin{eqnarray}
E_{gr} &=& \hbar \omega (1 + \gamma) 
\nonumber \\
\psi_{gr}(x) &=& Z_{0}^{1/2} ~ x^{1/2 + \gamma} ~
\exp[-\frac{m \omega}{2 \hbar} x^{2} ] ~ .
\end{eqnarray}
This is, of course, identical with the direct solution from the Schr\"odinger equation. The wave function is shown in Fig.[\ref{Fig_PotWave}].

\begin{figure}[thb]
\vspace{9pt}
\begin{center}
\includegraphics[scale=0.4,angle=270]{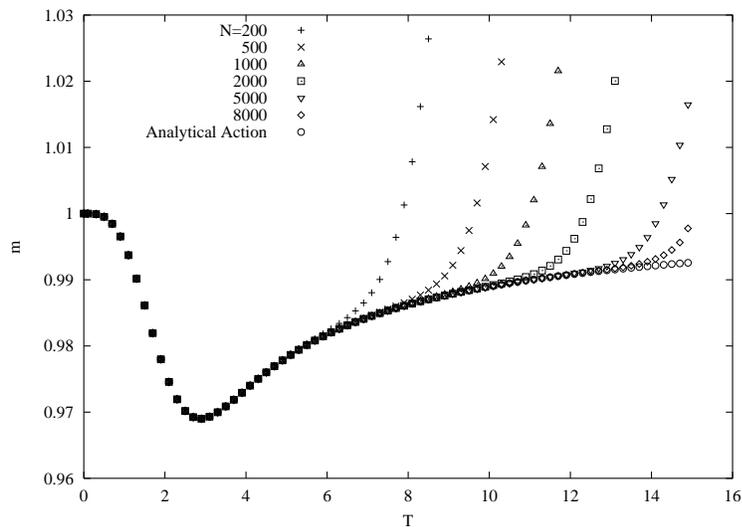}
\end{center}
\caption{Inverse square potential. Mass $\tilde{m}$ vs. transition time $T$. 
Dependence on the density of meshpoints $N$ per unit interval of $T$. }
\label{Fig_enmsh_m}
\end{figure}

\begin{figure}[thb]
\vspace{9pt}
\begin{center}
\includegraphics[scale=0.4,angle=270]{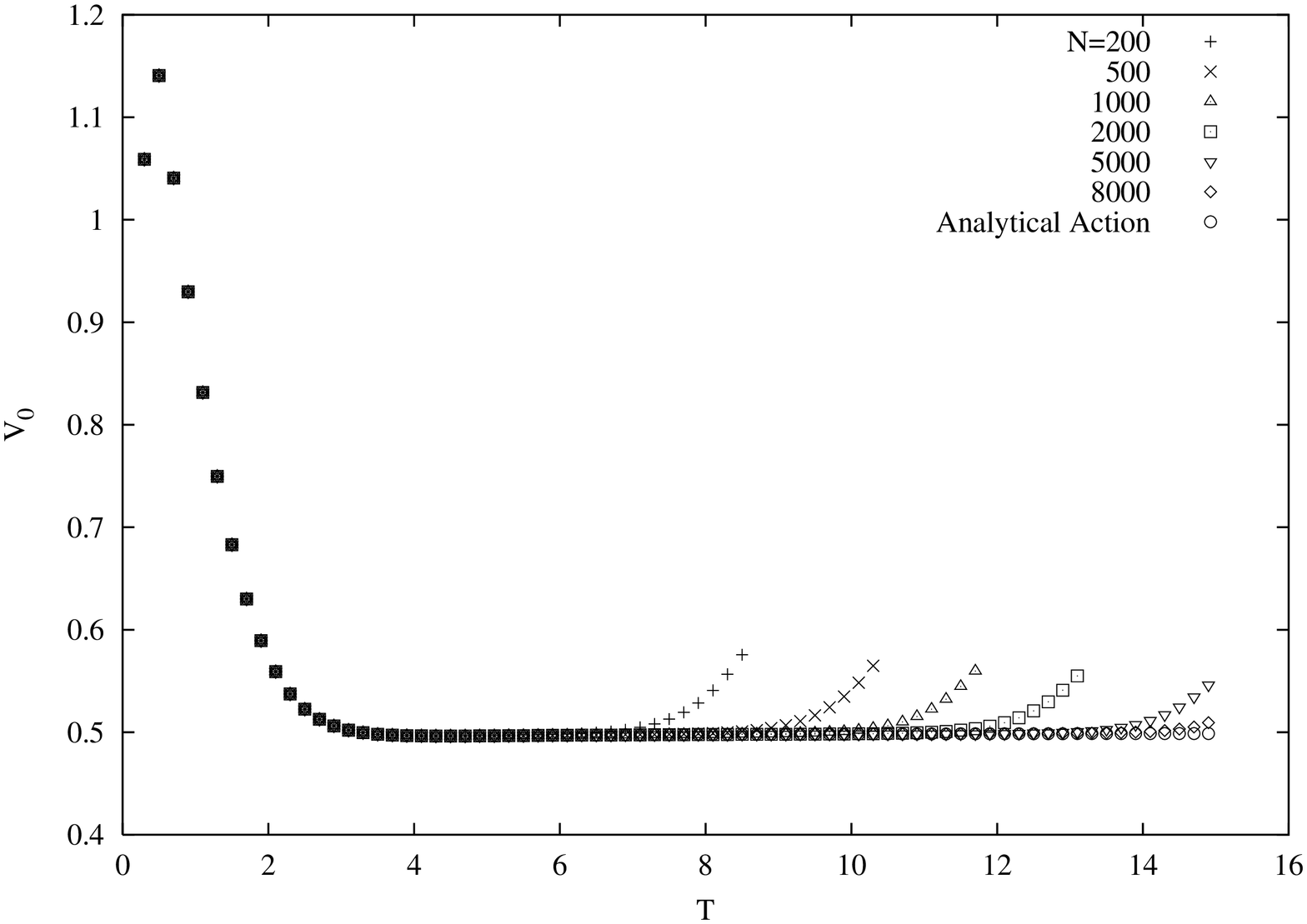}
\end{center}
\caption{Same as Fig.[\ref{Fig_enmsh_m}] for $\tilde{v}_{0}$. }
\label{Fig_enmsh_v0}
\end{figure}

\begin{figure}[thb]
\vspace{9pt}
\begin{center}
\includegraphics[scale=0.4,angle=270]{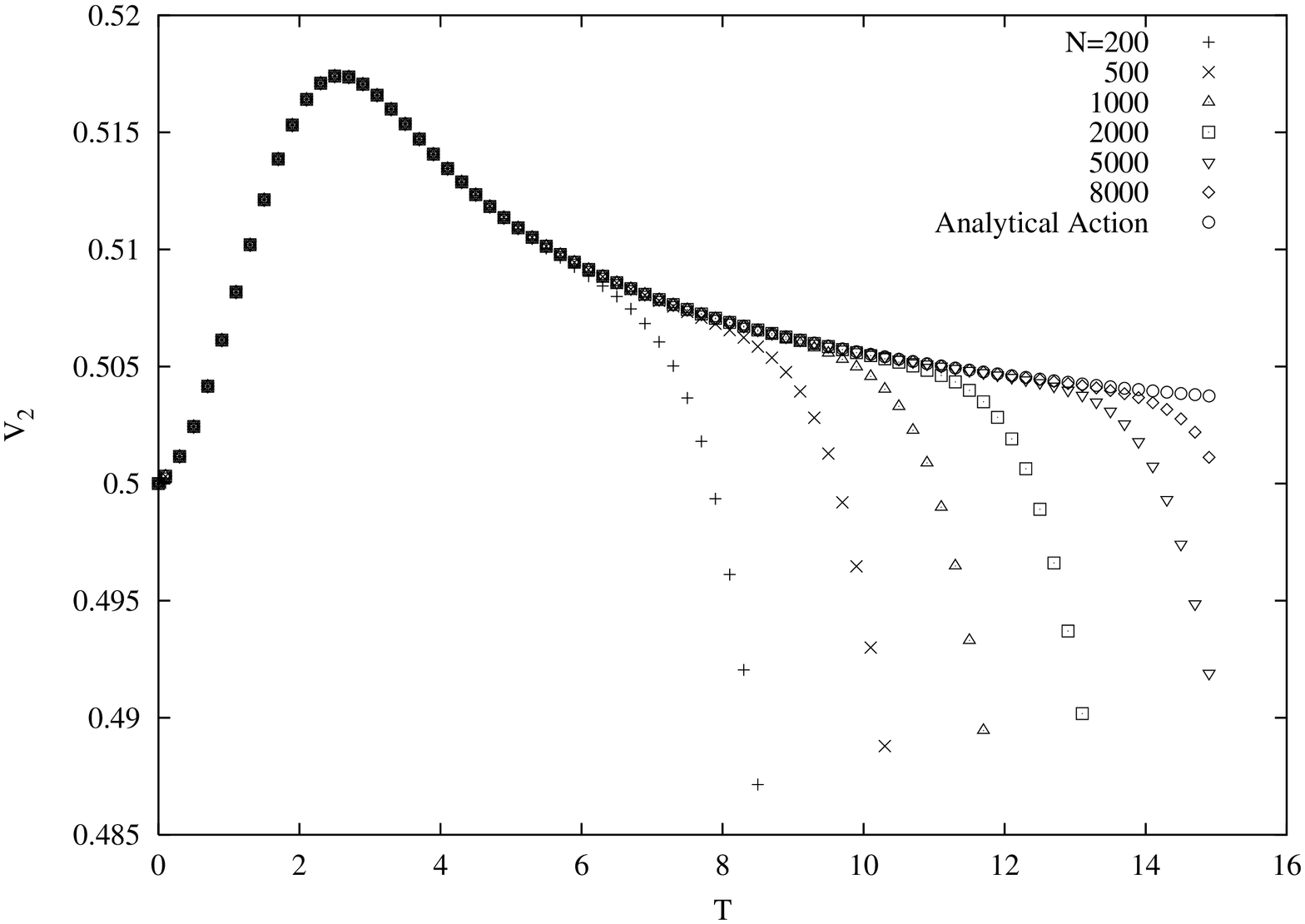}
\end{center}
\caption{Same as Fig.[\ref{Fig_enmsh_m}] for $\tilde{v}_{2}$. }
\label{Fig_enmsh_v2}
\end{figure}

\begin{figure}[thb]
\vspace{9pt}
\begin{center}
\includegraphics[scale=0.4,angle=270]{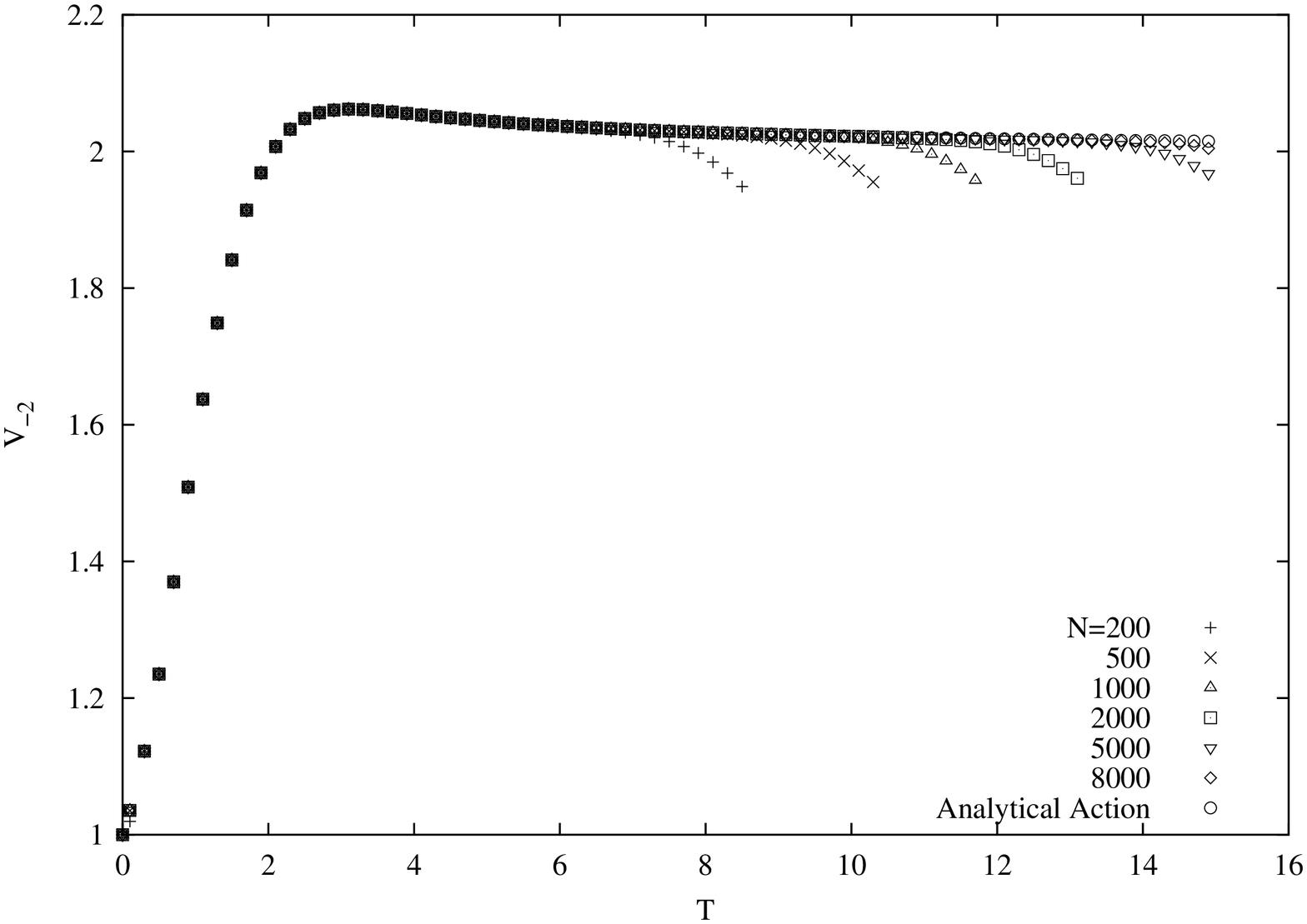}
\end{center}
\caption{Same as Fig.[\ref{Fig_enmsh_m}] for $\tilde{v}_{-2}$. }
\label{Fig_enmsh_v-2}
\end{figure}

\subsection{Dynamical time and length scales}
\label{sec:DynScal}

In this work we do a numerical study of a model to test the quantum action.
Our choice of the model has been influenced by the analytical solvability and less by the question if it plays a role in nature.
Consequently, absolute values in physical units of the model parameters (mass, potential parameters) are not of primary importance. We have expressed those parameters in dimensionless units. 
However, what is important are time and length scales, 
which are dynamically generated by the model, say a time scale $T_{sc}$ and a length scale $\Lambda_{sc}$. Those scales serve as reference values
to give a sense to statements like "for large transition times $T$", or a "small spatial resolution $\Delta x$", which means $T/T_{sc} >> 1$ and 
$\Delta x/\Lambda_{sc} << 1$, respectively. How to choose those dynamical scales? We have used as time scale
\begin{equation}
T_{sc} = \frac{1}{E_{gr}} ~ .
\end{equation}
As a length scale one may introduce the analogue of the Bohr radius of the ground state wave function. Another possibility is to define a length scale $\Lambda_{sc}$ by
\be
\int_{0}^{\Lambda_{sc}} dx ~ |\psi_{gr}(x)|^{2} = 0.95 ~ ,
\end{equation}
i.e. the length which covers $95\%$ of the probability of the ground state wave function. For the classical action considered in Eq.[\ref{eq:ClassStandParam}] of sect.[\ref{sec:GlobalFit}], those scale parameters are
\begin{eqnarray} 
&& T_{sc} = 0.4
\nonumber \\
&& \Lambda_{sc} \approx 2.35 ~ .
\end{eqnarray}

\subsection{Analytical results for the quantum action in the asymptotic regime}
\label{sec:AnalytResult}

In Ref.\cite{Q5} we have shown for the Euclidean asymptotic regime $T \to \infty$, that the following analytic relations exist between ground state energy $E_{gr}$, ground state wave function $\psi_{gr}$ and the quantum action.
\be
\label{GroundStateEner}
\tilde{V}_{min} = E_{gr} ~ ,
\end{equation}
i.e. the minimum of the quantum potential gives the ground state energy.  
Next, the ground state wave function can be expressed in terms of the parameters of the quantum action 
\be
\label{GroundStateWave}
\psi_{gr}(x) = \frac{1}{N} ~ e^{ - \int_{\tilde{x}_{min}}^{x} dx' ~ 
\sqrt{2 \tilde{m}( \tilde{V}(x') - \tilde{V}_{min} ) }/\hbar } ~ .
\end{equation}
Finally, there is a relation between classical and quantum action (transformation law) which reads (for $x > x_{min}$)
\be
\label{TransformLaw}
2 m(V(x) - E_{gr}) =  
2 \tilde{m}(\tilde{V}(x) - \tilde{V}_{min}) 
- \frac{\hbar}{2} \frac{ \frac{d}{dx} 2 \tilde{m} (\tilde{V}(x) - \tilde{V}_{min})}
{ \sqrt{2 \tilde{m}( \tilde{V}(x) - \tilde{V}_{min} ) } } ~
\mbox{sgn}(x-\tilde{x}_{min}) ~ .
\end{equation}
Here $\tilde{x}_{min}$ denotes the position of the minimum of the quantum potential.
Let us make an ansatz for the quantum action, characterized by a mass $\tilde{m}$ 
and a quantum potential of the form
\be
\label{QuantPot}
\tilde{V}(x) = \tilde{v}_{2} x^{2} + \tilde{v}_{-2} x^{-2} + \tilde{v}_{0} ~ ,
\end{equation}
and see if it satisfies transformation law, Eq.(\ref{TransformLaw}).
Using Eqs.(\ref{ClassPot},\ref{QuantPot}), inserting it into Eq.(\ref{TransformLaw}), and comparing the coefficients of the terms $x^{2}$, $x^{-2}$ and $x^{0}$, one obtains
\begin{eqnarray}
&& 2 \tilde{m} ~ \tilde{v}_{2} = m^{2} ~ \omega^{2} 
\nonumber \\
&& 2 \tilde{m} ~ \tilde{v}_{-2} - \hbar \sqrt{2 \tilde{m} ~ \tilde{v}_{-2} } 
= 2 m ~ g 
\nonumber \\
&& 4 \tilde{m} \sqrt{ \tilde{v}_{2} ~ \tilde{v}_{-2} }
+ \hbar \sqrt{2 \tilde{m} ~ \tilde{v}_{2}} = 2 m ~ E_{gr} ~ .
\end{eqnarray}
Those equations are equivalent to
\begin{eqnarray}
\label{Asymptotic}
&& \tilde{m} ~ \tilde{v}_{2} = \frac{1}{2} m^{2} ~ \omega^{2} 
\nonumber \\
&& \tilde{m} ~ \tilde{v}_{-2} = \frac{1}{2} \hbar^{2} ~ [\frac{1}{2} + \gamma]^{2}
\nonumber \\
&& E_{gr} = \hbar ~ [1 + \gamma] ~ .
\end{eqnarray}
For the case when the classical potential is given by the parameters $m=1$, $\hbar=1$, $\omega=1$ and $g=5$, this yields
\begin{eqnarray}
&& \tilde{m} ~ \tilde{v}_{-2} \to 0.5  
\nonumber \\
&& \tilde{m} ~ \tilde{v}_{2} \to 6.8507 ~ . 
\end{eqnarray}
The numerical solution, shown in Fig.[2]
gives 
\begin{eqnarray}
&& \tilde{m} ~ \tilde{v}_{-2} \to 0.500 
\nonumber \\
&& \tilde{m} ~ \tilde{v}_{2} \to 6.86 ~ . 
\end{eqnarray}
We observe that analytical and numerical results agree well.

Finally, let us compare the analytic behavior of the classical function 
$m V(x)$ (mass $\times$ potential) with the corresponding quantum function $\tilde{m} \tilde{V}(x)$. In particular, let us look at the term $x^{-2}$ which is singular at the origin. The corresponding term in the classical potential is 
$m g / x^{2}$ (Eq.(\ref{ClassPot}). Let us see what happens 
when $g \to 0$.
The classical term $m g / x^{2}$ is singular as a function of $x$ for any 
$g > 0$. For $g=0$ the singular term vanishes and the classical potential is regular as a function of $x$. This means that the classical potential has a singular behavior as a function of $g$ when $g \to 0$. 
Now let us look at the corresponding term in the quantum potential
Eq.(\ref{Asymptotic}) tells us
\be
\tilde{m} ~ \tilde{v}_{-2} = \frac{1}{2} \hbar^{2} ~ [\frac{1}{2} + \gamma]^{2}
= \frac{1}{2} \hbar^{2} \left[ 1 + \frac{4mg}{\hbar^{2}} \right] + O(g^{2}) ~ .
\end{equation}
This function has a regular behavior when $g \to 0$. 
Thus, in the limit $g \to 0$, the quantum potential is smoother than the 
classical potential. A similar behavior has been observed previously in a study
comparing a classical double well potential with the corresponding quantum double well potential and its corresponding instanton solutions \cite{Q2}. 
The quantum double well potential was found to have a lower barrier and the quantum instantons were found to be softer than their classical counterparts.

\section{Numerical results from global fit}
\label{sec:GlobalFit}

Let us first consider the classical action with the following parameters
\be
\label{eq:ClassStandParam}
m=1 ~, ~~~ v_{2} = 0.5 ~ , ~~~ v_{-2} = 1 ~ .
\end{equation}
This gives the ground state energy  
\be
E_{gr}=2.5 ~ .
\end{equation}
For the quantum action the following ansatz has been made,
\begin{eqnarray}
\tilde{S} &=& \int dt ~ \frac{1}{2} \tilde{m} \dot{x}^{2} + \tilde{V}(x) 
\nonumber \\
\tilde{V}(x) &=& \tilde{v}_{0} + \tilde{v}_{2} x^{2} + \tilde{v}_{-2} x^{-2}
+ \tilde{v}_{4} x^{4} + \tilde{v}_{-4} x^{-4} ~ .
\end{eqnarray}
The method to determine the parameters of the quantum action from given transitions amplitudes $G(b,T;a,0)$ proceeds by selecting a number of initial positions $x_{in}^{(\sigma)}$ and also final positions $x_{fi}^{(\rho)}$.
Making a guess for the parameters of the quantum action, one computes the trajectory corresponding to the quantum action with the boundary condition $x(t=0) = x_{in}^{(\sigma)}$ and $x(t=T)=x_{fi}^{(\rho)}$. Then one computes the value of the quantum action for this trajectory, say 
$\Sigma_{\rho \sigma}$. This is done for all $\sigma$ and $\rho$.
The goal is to make the error of the fit
\be
\label{eq:ErrG}
\sum_{\sigma,\rho} \left| G(x_{fi}^{(\rho}),T;x_{in}^{(\sigma)},0) - \exp[-\Sigma_{\rho \sigma}] \right|
\end{equation}
globally small. The parameters of the quantum action are those which minimize the error. For more details on the method see Ref.\cite{Q1}.    
In the calculations reported here we have used the following initial and final data: There are 2 initial points $x_{in}$ located in the interval $[4, 5]$.
There are 10 final points $x_{fi}$ equidistantly distributed in the interval $[0.5, 3]$. For the computation of the trajectories, the following resolution of time has been used: $N_{t}=500$ meshpoints for a time interval $\Delta T=1$.
For larger time intervals, the number of meshpoints has been increased proportionally.
An overview of the behavior of the parameters of the quantum action is displayed
in Figs.[\ref{Fig_e44_m} - \ref{Fig_e44_v-4}].
Fig.[\ref{Fig_e44_m}] shows the  mass of the quantum action $\tilde{m}$ as a function of transition time, which displays a smooth behavior. 
The behavior of the estimated errors seems to be smooth between $T=0$ up to $T=4$. At $T=4$, the error bars behave erratically.  
The behavior of the parameters $\tilde{v}_{0}$, $\tilde{v}_{2}$ etc. also looks smooth. In this calculation we included terms $x^{4}$ and $x^{-4}$ in the quantum potential. The data show that the coeffients are quite small and almost compatible with the value zero.
Fig.[\ref{Fig_e44_S}] presents the relative error of the quantum action vs. transition time $T$. 
The error starts out at $T=0$ from a value close to zero, then
increases and reaches at $T=1.5$ a maximum of $5 ~ 10^{-4}$ and finally decreases until $T=5$.
At $T=5$ there is a cusp.  
By adding terms $x^{-4}$ and $x^{4}$ to the quantum potential gives a qualitatively similar behavior. It reduces the maximum of the error by a factor of about 2.
Note that the parameters are shown on a linear scale while the error of the quantum action is shown on a logarithmic scale in order to resolve smaller numbers.
In order to see if higher order terms play a role in the quantum potential, we have carried out a calculation including the terms $x^{6}$ and $x^{-6}$ instead of $x^{4}$ and $x^{-4}$. The results (not shown here) are quite similar to the previous ones.

\bigskip

{\it Dependence on boundary points}. Next we look at the dependence of the quantum action parameters on the boundary points used in the fit. We have studied this in two ways:
(a) We kept the initial points $x_{i}$ fixed and varied the final points $x_{f}$.
(b) We kept the final points $x_{f}$ fixed and varied the initial points $x_{i}$.
In the first case, we have considered one initial point $x_{in}=0.3$ (kept fixed). 
As final points $x_{fi}$ we took 100 points uniformly distributed in an interval and varied those intervals $[2,3]$, $[5,6]$, $[9,10]$ and $[2,10]$.
The results are shown in Figs.[\ref{Fig_eevh_m} - \ref{Fig_eevh_v-2}]. 
One observes for all parameters of the quantum action that in the regime 
$T > 2 ~ (T > 5 T_{sc})$  there is no dependence on the location of the final boundary points. However, for $T < 2 ~ (T < 5 T_{sc})$ there is a noticeable dependence.

In the second case, as final boundary points $x_{f}$ we took 100 points uniformly distributed in the interval $[2,3]$ (interval kept fixed).
As initial points $x_{i}$ we took one point and varied it between 0.1 to 0.5.
The results are shown in Figs.[\ref{Fig_eevh23_m} - \ref{Fig_eevh23_v-2}].
Those results are qualitatively the same as in the first case.
Fig.[\ref{Fig_eevh23_m}] shows that the quantum mass depends on the boundary points for $T < 2 ~ (T < 5 T_{sc})$ and becomes independent for 
$T > 2 ~ (T > 5 T_{sc})$. The behavior for the parameters of the quantum potential is similar, 
however, the independence sets in a bit earlier at about 
$T \approx 1.6 = 4 T_{sc}$.

One may ask: What is the reason for such dependence on boundary points? In our opinion the most likely explanation is that the quantum action is not an exact parametrisation of the transition amplitude in this regime. Hence the parametrisation depends on the parameters of the fit.

The two previous cases represent a strong imbalance between the number of initial versus final boundary points. As a third case we have considered a more balanced selection of bondary points. We took 10 initial points, uniformly distributed in the interval $[1.5, 2.5]$ and 10 final points, uniformly distributed in the interval $[1.1, 2.1]$. The parameters of the quantum action are shown in Figs.[\ref{Fig_estand_m} - \ref{Fig_estand_mv-2}].  
The global relative error in fitting the transition matrix elements is shown in Fig.[\ref{Fig_estand_G}].

\bigskip

{\it Dependence on temporal resolution}. 
The determination of the parameters of the quantum action by a global best fit, Eq.(\ref{eq:ErrG}), requires to solve the equation of motion from $\tilde{S}$ for all pairs of boundary points and to compute the value $\tilde{\Sigma}$ of the quantum action along such trajectory. Such numerical calculations depend on the temporal resolution $\Delta t$. We have studied the convergence of the parameters of the quantum action as a function of the density of meshpoints.
We parametrize the density of meshpoints by taking $N_{t}$ meshpoints per unit time interval $\Delta T=1$. We have varied $N_{t}$ from 200 up to 8000.
The results are shown in Figs.[\ref{Fig_enmsh_m} - \ref{Fig_enmsh_v-2}].   
Fig.[\ref{Fig_enmsh_m}] shows for $T < 6 \approx 15 T_{sc}$ that a low density 
$N_{t}=200$ is sufficient to reach convergence. However, for larger $T$, the results diverge. If one desires convergence for, say $T=8 \approx 20 T_{sc}$, one needs to double the density of meshpoints. Such behavior persists if we want to maintain convergence for even larger $T$. E.g., for $T=14 \approx 35 T_{sc}$, we need a 20-fold higher density of meshpoints ($N_{t}=8000$).
Such exponential increase in the density of meshpoints could signal chaos caused by  imperfect numerical solutions, finite internal precision, rounding errors etc.

\bigskip

{\it Dependence on internal precision}. 
We have looked at the parameters of the quantum action in the asymptotic regime ($T \to \infty$) and attempted to analyze to role of internal computing precision. We performed a computation using FORTRAN giving approximately 
20 significant digits and a computation using MAPLE giving approximately 
30 significant digits. 
We found (not shown here) at $T \approx 15 \approx 38 T_{sc}$
a bifurcation between the computation with 20 and 30 digits. Moreover, 
the computation with 20 digits becomes flat for $T > 15$, while the computation with 30 digits becomes flat for $T > 19$. Thus, a tiny change in the internal precision by $\epsilon = 10^{-30}$ results in a change of $\tilde{m}$ at $T=19$ by a margin of $2.5 \%$. This signals a great sensibility to internal precision. Again, this could be interpreted as the presence of chaos, 
due to mathematical algoritms. (One should note that the 1-D Hamiltonian system under consideration is integrable and hence physical chaos is absent).

\begin{figure}[thb]
\vspace{9pt}
\begin{center}
\includegraphics[scale=0.4,angle=0]{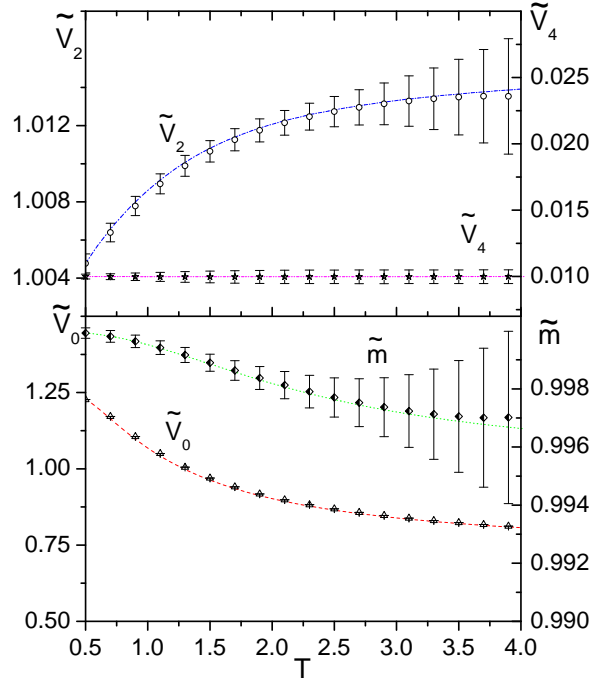}
\end{center}
\caption{Quantum action parameters for quartic potential. Comparison of results from flow equation (lines) vs. fit to transition amplitudes (symbols). }
\label{Fig_FluxFit}
\end{figure}

\section{Renormalisation group equation for temperature dependence 
of action parameters}
\label{sec:RenormGroup}

A global fit method has been used in sect.\ref{sec:GlobalFit} to determine the parameters of the quantum action. 
The method is quite costly from the computational point of view. It requires
first to compute the quantum transition matrix elements and then to carry out a multi-parameter fit. 
Here we want to suggest an alternative non-perturbative method. It is inspired by the idea of the renormalisation group equation in quantum field theory.
In $QFT$ the dependence of action parameters upon variation of a scale parameter (cut-off $\Lambda$, lattice spacing $a_{s}$ and $a_{t}$) is governed by a renormalisation group equation (e.g. Callan-Symanzik).
In the Q.M. system considered here, we are close to the 
continuum limit (in the example considered below $\Delta x/a_{B} = 0.045$ and $\Delta t/a_{B}=5 \times 10^{-6}$, where $a_{B}$ is the ground state Bohr radius of the ground state wave function). 

Here we draw a parallel between the scale-dependence of an observable action parameter in field theory (e.g. coupling $g$ in $QCD$) and the dependence 
of the parameters of the quantum action upon (Euclidean) transition time $T$.
Finite temperature physics is obtained from transition amplitudes in Euclidean transition time $T$, and the inverse temperature $\beta$ is related to $T$ via
$\beta = T/\hbar$.
As scale and temperature dependence of the action are similar, we apply 
here the term renormalisation group to describe temperature dependence
of the quantum action \cite{Q3}.
As a result we will end up with a differential equation (flow equation) for the 
parameters of the quantum action. In contrast to the global fit method, this flow equation does not require as input the transition amplitudes. 
It does, however, require initial data, i.e. the parameters of the quantum action at some initial temperature.

We consider the transition amplitude as a function of 
$x$ and $t$, keeping initial data $x_{in}$, $t_{in}$ fixed. It satisfies the Schr\"odinger equation 
\begin{equation}
\label{eq:SchrodEq}
- \hbar \frac{d}{d t} G(x,t;x_{in},t_{in})  
= \left[ - \frac{\hbar^{2}}{2 m} 
\frac{d^{2}}{dx^{2}} + V(x) \right] G(x,t;x_{in},t_{in}) ~ ,
\end{equation}
with the initial condition
\be
\label{eq:InitCond}
\lim_{t \to t_{in}} G(x,t;x_{in},t_{in}) = \delta(x - x_{in}) ~ .
\end{equation}
In the limit $t \to t_{in}$, the transition amplitude 
is given by the classical action,
consistent with the initial condition, Eq.(\ref{eq:SchrodEq}),
\begin{equation}
\label{eq:ShortTime}
\lim_{t \to t_{in}} \left\{ 
G(x,t;x_{in},t_{in}) - 
\frac{1}{Z} \exp[ -\frac{1}{\hbar} \Sigma_{cl}|_{x_{in},t_{in}}^{x,t} ] 
\right\}
\to 0 ~ .
\end{equation}
Going over to inverse temperature $\beta$, the parametrisation of (Euclidean) 
transition amplitudes in terms of the (Euclidean) quantum action reads
\be
\label{eq:QActionBeta}
G_{\beta} = \tilde{Z}_{\beta} \exp[ - \tilde{\Sigma}_{\beta} ] ~ .
\end{equation}
Combining Eq.(\ref{eq:QActionBeta}) with Eq.(\ref{eq:SchrodEq}) (expressed in terms of $\beta$) implies   
\be
\label{SchrodEqSigma}
- \frac{1}{\tilde{Z}_{\beta}} \frac{d \tilde{Z}_{\beta}}{d \beta} 
+ \frac{d \tilde{\Sigma}_{\beta} }{ d\beta } + \frac{\hbar^{2}}{2m} 
[ (\frac{d \tilde{\Sigma}_{\beta}}{dx})^{2} - \frac{d^{2} \tilde{\Sigma}_{\beta}}{dx^{2}} ] 
- V = 0 ~ .
\end{equation}
$\tilde{\Sigma}_{\beta}$ is given by the quantum action along its classical trajectory from $x_{in}, \beta_{in}=0$ to $x, \beta$
\begin{eqnarray}
\label{SigmaParam}
\tilde{\Sigma}_{\beta} &=& \tilde{S}_{\beta}[\tilde{x}_{cl}]|_{x_{in},0}^{x,\beta}
= \int_{x_{in},0}^{x,\beta} d\beta' ~ \frac{\tilde{m}}{2 \hbar^{2}} 
(\frac{ d \tilde{x}_{cl}}{d \beta'})^{2} + \tilde{V}(\tilde{x}_{cl}) 
\nonumber \\
&=&  \tilde{\Sigma}_{\beta} \left[ \tilde{m}(\beta),\tilde{v}_{0}(\beta),\tilde{v}_{1}(\beta),\dots,
x,\beta \right] ~ . 
\end{eqnarray} 
Here $\tilde{v}_{k}(\beta), ~ k=0,1,2,\dots$ denotes a set of parameters of the quantum potential (as an example $\tilde{v}_{k}(\beta)$ may be the coefficients of the quantum potential $\tilde{V}(x,\beta)$ expanded in terms of polynomials).
In general the number of terms is infinite. The weight of higher terms decreases rapidly. Eqs.(\ref{SigmaParam},\ref{SchrodEqSigma}) yield the  renormalisation group equation for the action parameters, 
$\tilde{m}(\beta), \tilde{v}_{0}(\beta),\tilde{v}_{1}(\beta),\dots$
as function of $\beta$,
\be
\label{RenormGroup}
- \frac{1}{\tilde{Z}_{\beta}} \frac{d \tilde{Z}_{\beta}}{d \beta} 
+  
\frac{\partial \tilde{\Sigma}_{\beta}}{\partial \tilde{m}} 
\frac{\partial \tilde{m}}{\partial \beta } +
\sum_{k} \frac{\partial \tilde{\Sigma}_{\beta}}{\partial \tilde{v}_{k}} 
\frac{\partial \tilde{v}_{k}}{\partial \beta } +
\frac{\partial \tilde{\Sigma}_{\beta}}{\partial \beta} 
+ \frac{\hbar^{2}}{2m} [ (\frac{d \tilde{\Sigma}_{\beta}}{dx} )^{2} 
- \frac{d^{2} \tilde{\Sigma}_{\beta}}{dx^{2}} ] - V = 0 .
\end{equation}
This equation is valid for all $x$, $x_{in}$. The parameters $\tilde{m}(\beta)$, $\tilde{v}_{k}(\beta)$ are independent of $x$, $x_{in}$. This constitutes a system of equations to determine 
$\partial \tilde{Z}_{\beta}/\partial \beta$,
$\partial \tilde{m}/\partial \beta$ and $\partial \tilde{v}_{k}/\partial \beta$.
The solution of the differential equations requires 
initial values. 
We noted the property that the quantum action goes over to the classical action in the limit when the transition time goes to zero. This suggests to take for $\beta=0$ as initial values 
$\tilde{Z}_{\beta}(\beta \to 0) \sim Z(\beta \to 0)$ (note: singularity at origin), $\tilde{m}(\beta=0) = m$, 
$\tilde{v}_{k}(\beta=0) = v_{k}, ~ k=0,1,\dots$.
Starting from initial values the renormalisation group equation determines the flow of the renormalized parameters when $\beta$ increases.

\begin{figure}[thb]
\vspace{9pt}
\begin{center}
\includegraphics[scale=0.4,angle=270]{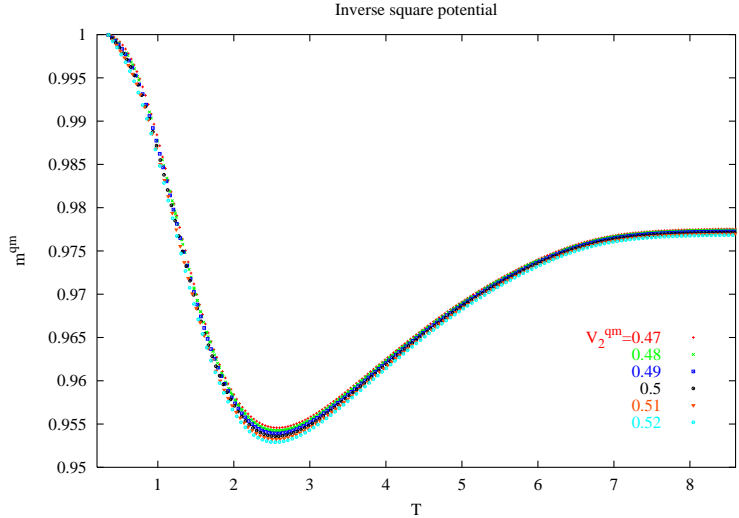}
\end{center}
\caption{Inverse square potential. Quantum action parameters obtained from flow equation. Dependence on initial data of flow equation. 
Mass $\tilde{m}$ vs. transition time $T$. }
\label{Fig_diffconv_m}
\end{figure}

\begin{figure}[thb]
\vspace{9pt}
\begin{center}
\includegraphics[scale=0.4,angle=270]{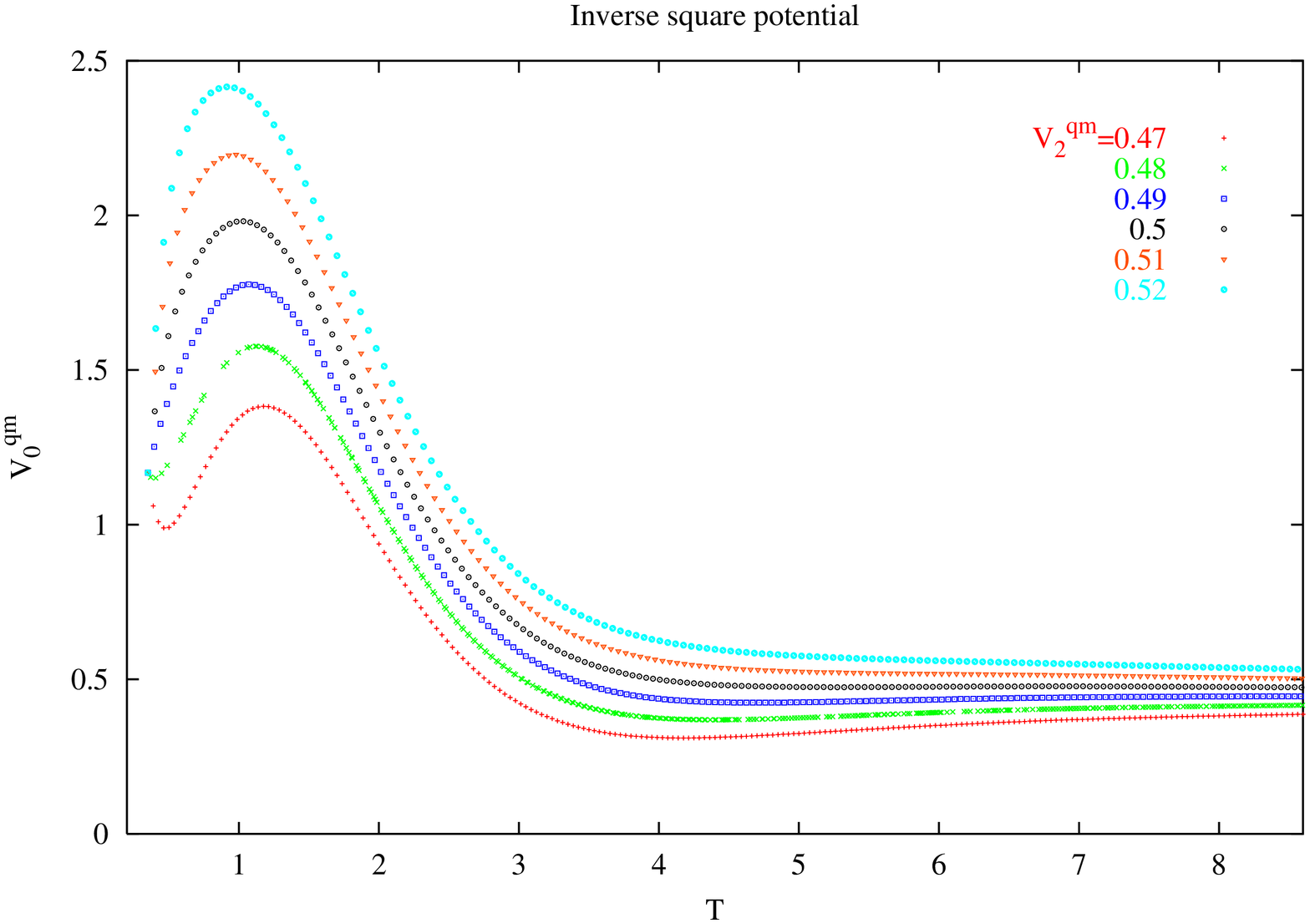}
\end{center}
\caption{Same as Fig.[\ref{Fig_diffconv_m}] for $\tilde{v}_{0}$. }
\label{Fig_diffconv_v0}
\end{figure}

\begin{figure}[thb]
\vspace{9pt}
\begin{center}
\includegraphics[scale=0.4,angle=270]{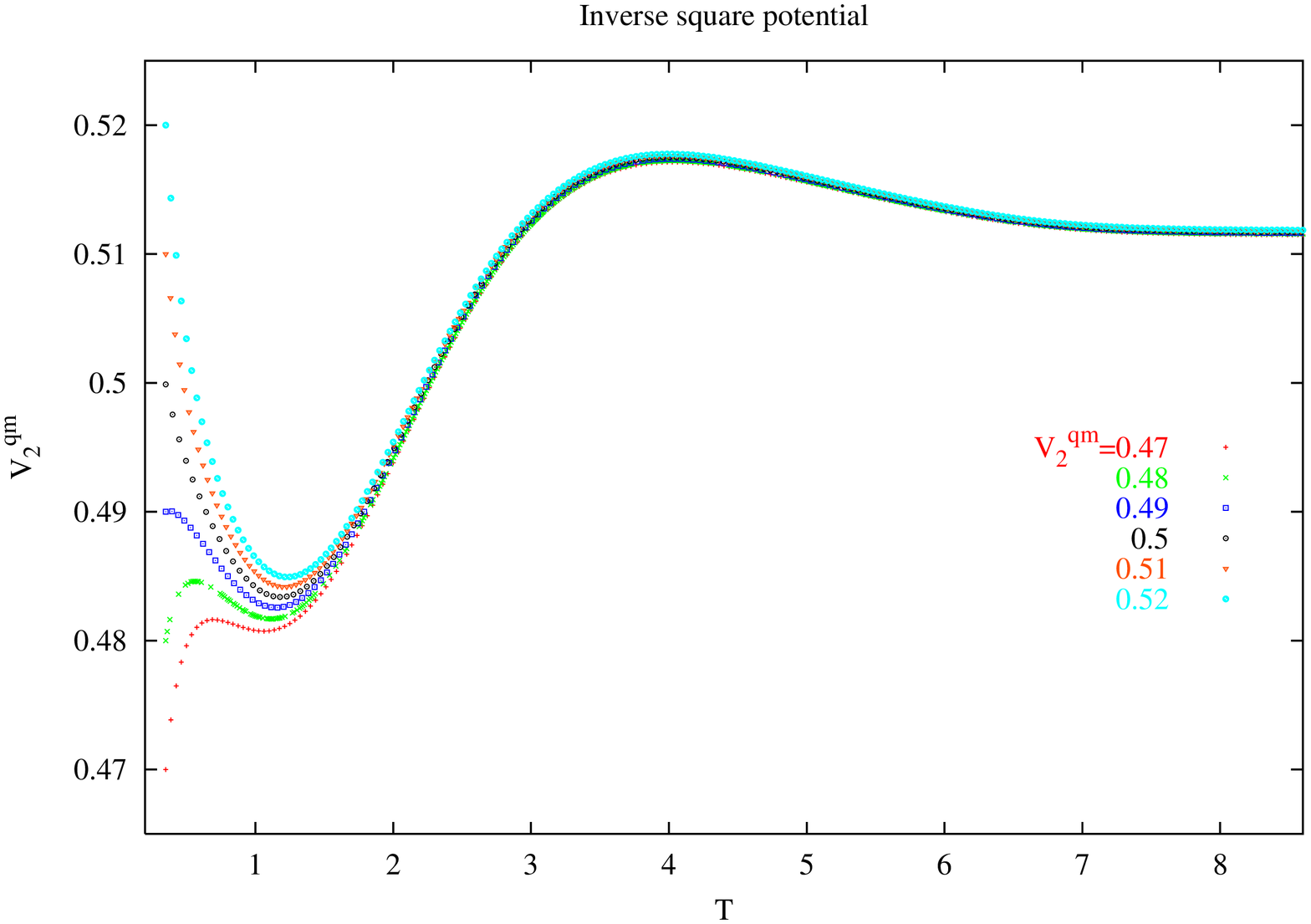}
\end{center}
\caption{Same as Fig.[\ref{Fig_diffconv_m}] for $\tilde{v}_{2}$. }
\label{Fig_diffconv_v2}
\end{figure}

\begin{figure}[thb]
\vspace{9pt}
\begin{center}
\includegraphics[scale=0.4,angle=270]{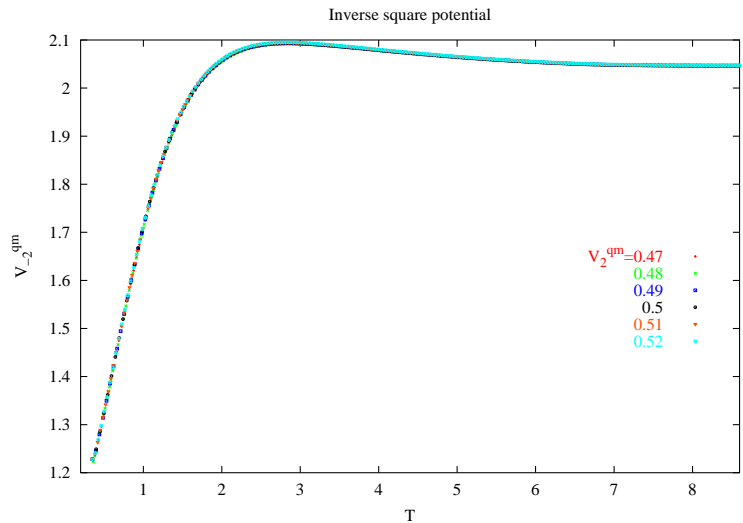}
\end{center}
\caption{Same as Fig.[\ref{Fig_diffconv_m}] for $\tilde{v}_{-2}$. }
\label{Fig_diffconv_v-2}
\end{figure}

{\it Flow equation for quartic potential}.
As a first test of the method we have applied it to the quartic potential in 1-D, given by the classical action
\begin{eqnarray}
&& S = \int dt ~ \frac{1}{2} m \dot{x}^{2} + V(x)
\nonumber \\
&& V(x) = v_{2} x^{2} + v_{4} x^{4}
\nonumber \\
&& m = 1, ~ v_{2}=1, ~ v_{4}=0.01 ~ .
\end{eqnarray}
We have computed numerically the solution of the flow equation of the quantum action parameters, Eq.(\ref{RenormGroup}). We have taken a single initial boundary point $x_{in}$ and considered transitions to a number of final boundary points $x_{f}$. We have compared this with a global fit of the quantum action to $QM$ transition amplitudes (see Ref.\cite{Q1}). A comparison is shown in Fig.[\ref{Fig_FluxFit}]. One observes good agreement in the range 
$\beta = 0.5$ to $\beta = 4$.

\bigskip

{\it Flow equation for inverse square potential}.
We considered the action with the inverse square potential, 
with the parameters of the classical action given by
\begin{equation}
m = 1, ~ , v_{2} = 0.5, ~ v_{-2} = 1 ~ ,
\end{equation}
which are the same parameters as used in Eq.(\ref{eq:ClassStandParam}) of sect.(\ref{sec:GlobalFit}).
We looked at the behavior of the parameters of the quantum action obtained by solving the flow equation. 
First, we want to study the stability of the parameters of the quantum action
under variation of initial conditions.
In particular we chose the following initial conditions: At $T_{init}=0.35$, we used as initial data parameters obtained by the global fit method,
\begin{equation}
\tilde{m} = 0.99994533, ~
\tilde{v}_{0} = 1.1676274, ~
\tilde{v}_{-2} = 1.2280919 ~ .
\end{equation}
The inital value of $\tilde{v}_{2}$ has been varied between 0.47 to 0.52.
In particular, we have studied how the quantum action parameters depend on the choice of initial value of $\tilde{v}_{2}$.
As initial boundary points we used a single point, $x_{in}=10$.   
As final boundary points $x_{f}$ we used 30 points uniformly distributed 
in the interval $[0.2, 7]$. The differential equations have been solved with a resolution $\Delta \beta = 3.75 ~ 10^{-3}$.
The results are shown in Figs.[\ref{Fig_diffconv_m} - \ref{Fig_diffconv_v-2}]. 
One observes for $\tilde{m}$ and $\tilde{v}_{-2}$, 
shown in Figs.[\ref{Fig_diffconv_m}, \ref{Fig_diffconv_v-2}], 
a very weak influence of the variation of the initial value of $\tilde{v}_{2}$. 
However, in $\tilde{v}_{0}$, Fig.[\ref{Fig_diffconv_v0}], there is a substantial variation. It is interesting to note that $\tilde{v}_{2}$, shown in Fig.[\ref{Fig_diffconv_v2}], converges rapidly 
for $\beta > 2$. In other words, except for $\tilde{v}_{0}$, all parameters of the quantum action, although starting from different initial data, collapse to a single curve, when $\beta > 2$. The flow equation method shows stability of results under variation of initial data.

Second, we looked at the dependence of the solution on the location of final boundary points. In this case, we took at $T_{init}=0.35$ the following initial data
\begin{equation}
\tilde{m} = 0.99994533, ~
\tilde{v}_{0} = 1.1676300, ~
\tilde{v}_{2} = 0.4998810, ~
\tilde{v}_{-2} = 1.2280900 ~ .
\end{equation}
We took a single initial boundary point $x_{in}=10$,
As final boundary points $x_{f}$ we took 30 points uniformly distributed in some interval $[c,d]$, where $c=0.2$ and $d$ has been varied from $d=5$ to $d=17$.                The results are displayed in Figs.[\ref{Fig_eqdiff_mv-2} - \ref{Fig_eqdiff_Dmv-2}]. Fig.[\ref{Fig_eqdiff_mv-2}] shows a weak dependence 
on the location of final boundary points. There is some visible dependence for $T < 2$. For $T > 2$, this dependence is invisible from the figure. However, 
because we have an analytic prediction for $\tilde{m} \tilde{v}_{-2}$ in the limit $t \to \infty$, we can compare the numerical results with the analytic prediction. The difference is plotted in Fig.[\ref{Fig_eqdiff_Dmv-2}]. 
One observes that some dependendence on the location of the final boundary points continues to exists beyond $T = 2$ (note logarithmic scale). However, 
in the limit of large $T$, this difference tends to zero (within numerical errors).

\begin{figure}[thb]
\vspace{9pt}
\begin{center}
\includegraphics[scale=0.4,angle=270]{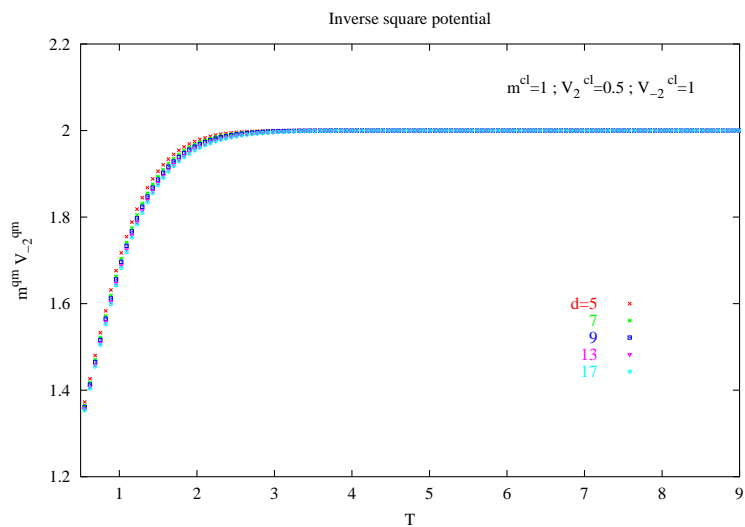}
\end{center}
\caption{Inverse square potential. Quantum action parameters obtained from flow equation. Dependence on location of interval $[c,d]$ of final boundary points.
$\tilde{m} \tilde{v}_{-2}$ vs. $T$. }
\label{Fig_eqdiff_mv-2}
\end{figure}

\begin{figure}[thb]
\vspace{9pt}
\begin{center}
\includegraphics[scale=0.4,angle=270]{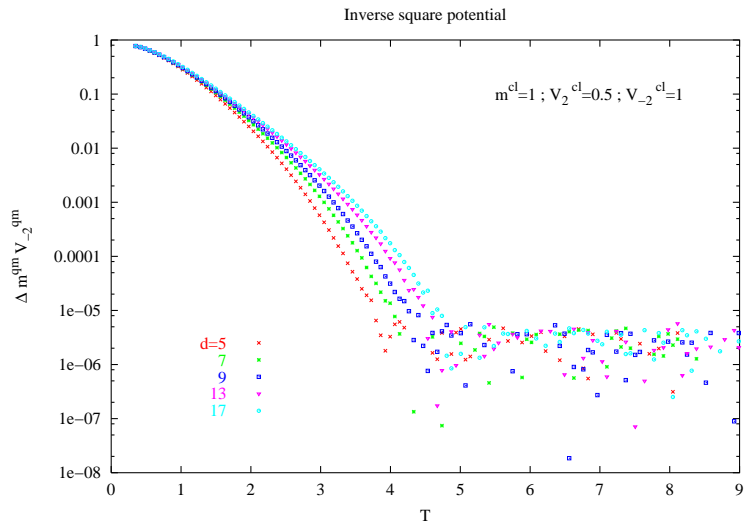}
\end{center}
\caption{Same as Fig.[\ref{Fig_eqdiff_mv-2}] for 
$\Delta \tilde{m} \tilde{v}_{-2}$. }
\label{Fig_eqdiff_Dmv-2}
\end{figure}

\section{Discussion}
\label{sec:Discuss}
We have studied a 1-D Hamiltonian system, which is integrable, given by an inverse square potential. It has the attractive feature that its 
quantum transition amplitudes are analytically known. We have chosen this system to test the validity of the quantum action. The knowledge of transition amplitudes eliminates one source of error in the construction of the quantum action. We carried out a numerical study and determined the parameters of the quantum action by two different methods: via global fit and via flow equation.

We consider the following observations as most important:
As a function of transition time $T$ (or inverse temperature $\beta$), the behavior of the parameters of the quantum action can be divided in three regimes: (i) Regime of small $T$, say $0 < T < T_{sc}$. For small $T$ high lying energies will contribute in thermodynamic observables. 
In atomic physics, the high-lying states of the hydrogen atom are known as Rydberg states and the wave functions are known to have a semi-classical behavior. Thus this regime is the semi-classical regime. This is consistent also with the well known property that quantum transition amplitudes in the limit 
$T \to 0$ are well described by $G = Z \exp[\frac{i}{\hbar} S_{cl}]$, i.e. 
the path integral of the transition amplitude is well approximated by the classical action evaluated along a single path - the trajectory of the classical action. It is interesting to note that trace formulas (Gutzwiller and extensions) work quite well in the semiclassical regime. Then the path integral can be well approximated by periodic orbits. 
In this work we observe that the quantum action also works well in the semiclassical regime. More precisely, it works the better the smaller the value of $T$ is. 

(ii) There is another regime, that of large $T$, say $5 T_{sc} < T < \infty$.
This regime is opposite of the semiclassical regime, i.e. the deep quantum regime. In thermodynamics, large $\beta$ means small temperature. In this limit the Feynman-Kac formula holds and the physics is described by the ground state properties. It is not known if the trace formulas work in this regime (the authors are not aware of any evidence).
It is known that the quantum action becomes an exact parametrisation in this limit \cite{Q4}. Here we have analyzed the behavior of the quantum action numerically. We found that both, the global fit method, as well as the flow equation method give results in agreement with each other and also with the analytic result. 

(iii) Finally, there is an intermediate regime, say $T_{sc} < T < 5 T_{sc}$.
This is the regime where the parametrisation of the transition amplitude gives the largest relative global errors (see Figs.[\ref{Fig_estand_G}, \ref{Fig_e44_S}]), also manifested by the strongest dependence on boundary points. Although this error is small, it is by several orders of magnitude larger than the error in the regime of large $T$. How can this be understood?
The following scenarios are possible:
First, the ansatz for the quantum potential may be incomplete.  
There may be a need to include local terms in the quantum potential beyond those occuring in the classical potential. We have carried out some steps in this direction. We included in the quantum potential the following terms, all absent from the classical potential: $x^{4}$ and $x^{-4}$. Some improvement has been found (see Fig.[\ref{Fig_e44_S}]), however, it is less than one order of magnitude.  
Similar results have been obtained when incorporating the terms $x^{6}$ and $x^{-6}$. Of course it is possible that there are terms missing of a form quite different from the polynomial form considered here. However, we consider that as unlikely. The results seem to indicate that this error is not mainly due to a lack of terms in the quantum potential.  

Second, the parametrisation of transition amplitudes by the quantum action may be incomplete in the sense that it is not sufficient to evaluate the quantum action  along its corresponding classical trajectory, which minimizes the quantum action, but other trajectories need to be taken into account.
Let us be more specific. For stiff potentials, like the quartic potentials, it is well known that there exists an infinity of classical 
trajectories, for a given pair of boundary points.
This has been discussed in some detail by Schulman \cite{Schulman,Lam67}.
Those trajectories are all solutions of the classical equations of motion, and all correspond to the same pair of boundary points, but they differ in the value of its action. On the other hand, we observed that the quantum action works well  in the limit of large $T$, i.e. the Feynman-Kac regime, where the physics is given by the ground state properties. When lowering $T$ then gradually the first excited state, the second excited state, etc. will contribute in the partition function and the (Euclidean) transition amplitudes. At the same time we notice 
an increase of the error in the parametrisation by the quantum action when lowering $T$.
We propose the following hypothes:
In a regime where higher lying states need to be taken into account in the transition amplitudes, then the parametrisation by the quantum action needs to take into account trajectories of the quantum action with higher values of the action. There is an interesting geometrical analogy between the wave functions of excited states and those "higher" trajectories:
The ground state wave function has no nodes, but the wave functions of excited states have nodes (its number increasing with energy).
Correspondingly "higher" trajectories have wiggles (its number increases with the value of the action).

Third, the concept of the quantum action as a means to parametrize transition amplitudes may be plain wrong. But then one has to ask: Why does the quantum action work in the regime of small and large $T$?
In our opinion the most promising route to further explore those questions 
is to carry out high precision numerical simulations in different models.

\vspace{0.5cm}
\noindent {\bf Acknowledgements} \\ 
H.K. and K.M. are grateful for support by NSERC Canada. G.M. and D.H. have been supported in part by FCAR Qu\'ebec. 
For discussions and constructive suggestions H.K. is very grateful to A. Okopinska and L.S. Schulman.

\end{document}